\documentclass[a4paper,11pt]{article}
\pdfoutput=1 

\usepackage{jheppub} 
\usepackage{graphicx}
\usepackage{amsmath}
\usepackage{slashed}
\usepackage{amssymb}
\usepackage[mathscr]{euscript}
\usepackage{enumerate}
\usepackage{hyperref}
\usepackage{subfigure}

\hypersetup{
	colorlinks=true,
	linktoc=page,
	linkcolor=blue,
	citecolor=blue,
	urlcolor=blue
	}

\newcommand{\hs}{\hspace{0.15mm}}

\newcommand{\ms}{\mathbb{S}}
\newcommand{\mv}{\mathbb{V}}
\newcommand{\mt}{\mathbb{T}}

\usepackage[T1]{fontenc} 

\title{On the stability of gravity with Dirichlet walls}

\author[a]{Tom\'as Andrade}
\author[b]{William R. Kelly}
\author[b]{Donald Marolf}
\author[c]{Jorge E. Santos}

\affiliation[a]{Rudolf Peierls Centre for Theoretical Physics \\ University of Oxford, 1 Keble Road, Oxford OX1 3NP, UK}
\affiliation[b]{University of California at Santa Barbara \\ Santa Barbara, CA 93106, USA}
\affiliation[c]{DAMTP, Centre for Mathematical Sciences, University of Cambridge, Wilberforce Road, Cambridge CB3 0WA, UK}

\emailAdd{tomas.andrade@physics.ox.ac.uk}
\emailAdd{wkelly@physics.ucsb.edu}
\emailAdd{marolf@physics.ucsb.edu}
\emailAdd{J.E.Santos@damtp.cam.ac.uk}

\abstract{Dirichlet walls -- timelike boundaries at finite distance from the bulk on which the induced metric is held fixed -- have been used to model AdS spacetimes with a finite cutoff.  In the context of gauge/gravity duality, such models are often described as dual to some novel UV-cufoff version of a corresponding CFT that maintains local Lorentz invariance.
We study linearized gravity in the presence of such a wall and find it to differ significantly from the seemingly-analogous case of Dirichlet boundary conditions for fields of spins zero and one.  In particular, using the Kodama-Ishibashi formalism, the boundary condition that must be imposed on scalar-sector master field with harmonic time dependence depends explicitly on their frequency.  That this feature first arises for spin-2 appears to be related to the second-order nature of the equations of motion.  It gives rise to a number of novel instabilities, though both global and planar Anti-de Sitter remain (linearly) stable in the presence of large-radius Dirichlet cutoffs.  The instabilities arise on the outside of spherical Dirichlet walls, and also inside sufficiently large such spherical walls in de Sitter space.     We analyze both inside and outside of flat and spherical walls in Minkowski, de Sitter, and anti-de Sitter space, as well as in certain black hole spacetimes and find stability for cases not mentioned above. In particular, we find no linear instabilities in the presence of flat walls.  We also find evidence supporting the conjecture that neutral black holes are repelled by Dirichlet walls.}

\begin{document}
\maketitle
\flushbottom

\section{Introduction}

It is often interesting to consider only a portion of a given spacetime, a construction that in field theory is sometimes known as ``putting the system in a box.'' See \cite{Hawking:1976de} for an early
application of this idea to gravitational physics.
The introduction of geometric cutoffs or hard walls in a given spacetime
is commonly used to model more realistic situations in which the wall, or rather, the boundary
conditions whose presence the wall requires, is the result of some physical property \cite{Witek:2010qc, Cardoso:2004nk,Herdeiro:2013pia}.
The point of doing so is to typically to make a mathematically simpler toy model of a complicated phenomenon.
Geometric cutoffs are also widely used in the
context of gauge/gravity duality, in which they not only play the role of dual ultraviolet \cite{Brattan:2011my, Heemskerk:2010hk,
Faulkner:2010jy, Marolf:2012dr, Bredberg:2010ky, Bredberg:2011jq} and infra-red cutoffs \cite{Hartman:2013qma},
but also provide toy models of physical boundaries in the dual gauge theories \cite{Takayanagi:2011zk}. (See \cite{Aharony:2010ay} for more exact descriptions.)  In the particular case where the intrinsic metric is fixed, we call this surface a ``Dirichlet wall.''

With such motivations in mind, we recently considered the problem of a black hole moving
through a spacetime with a Dirichlet wall \cite{Andrade2015moduli}. In order to make the problem
analytically tractable, we considered the extreme multi-black-hole solutions of \cite{Majumdar:1947eu, Papapetrou}.  Using the method of
images to satisfy the boundary condition at the wall, one may obtain geometries describing static black holes in the presence of a flat Dirichlet wall.  Their relative motion may then be studied perturbatively in a small velocity (or moduli space) approximation as in \cite{Ferrell:1987gf}.  Though our analysis raised questions as to whether this approximation is really well-defined, we found robust evidence that the kinetic energy of such extreme black holes can become {\it negative} near a Dirichlet wall.  This signals a likely dynamical instability.

However, \cite{Andrade2015moduli} found no evidence of instability for a purely gravitational theory in which all black holes are uncharged. To gain further insight into gravity with Dirichlet boundary conditions,  we now investigate the linear stability of vacuum gravity about various Dirichlet-wall backgrounds\footnote{To be precise, this manuscript will be entirely devoted to the study of linear-mode stability of Dirichlet-wall backgrounds.}.  We choose backgrounds constructed from the usual maximally-symmetric spacetimes (Minkowski, Anti-de Sitter (AdS), and de Sitter (dS)) in arbitrary numbers of dimensions by adding spherical or planar static Dirichlet walls.  The planar cases turn out to be stable.  In the spherical cases we can distinguish perturbations outside the wall from those inside, with the latter being confined to a spherical cavity.  In stark contrast to the familiar results for spins zero and one, the systems outside spherical cavities are all unstable.  Turing to the situation inside the cavity, the Minkowski and AdS systems inside spherical walls are stable though the corresponding dS system is unstable for sufficiently large spherical walls.  In some situations these linear stability results
can be established analytically. When this is not the case we resort to numerical computations.

For the  systems inside spherical walls (and inside a flat wall in AdS), we also study linear perturbations in the presence of black holes.  We find new modes that describe stable oscillations of the black hole within spherical cavities in either Minkowski or AdS space, thereby confirming the intuition that horizons promote stability. This result also supports
 the prediction of \cite{Andrade2015moduli} that uncharged black holes are repelled from Dirichlet walls.  We also find the expected instability to roll down the gravitational potential toward the cosmological horizon in de Sitter space -- provided the black hole begins far enough from the Dirichlet wall that the above repulsion can be neglected at short times.

Our analysis relies heavily on the Kodama-Ishibashi formalism of
\cite{Kodama2003}, which decouples the various types of gravitational perturbations (known as scalar, vector, and tensor) by writing them in terms of so-called master fields.  An important part of our work below is to determine how Dirichlet boundary conditions on the metric should be implemented in terms of the master fields.  The boundary conditions for vector and tensor master fields are standard ones (e.g. Dirichlet for the tensor modes), and as a result these modes are stable\footnote{The master field formalism has been used to show stability of
black hole solutions without Dirichlet walls in \cite{Kodama:2003kk,Ishibashi:2003ap,Kodama:2003ck,Kodama:2004gz}.
These works take the fluctuations (and hence the master fields) to have compact support, so that
they can show linear stability by constructing the appropriate self-adjoint operators whose eigenvalues are
the squared frequencies of the Fourier modes.  The Dirichlet boundary condition satisfied at the wall by our tensor-mode master fields often allows us to adapt this argument to show their stability.  But for the  vector and scalar master fields this argument can be obstructed  by the fact that do not vanish at the wall. We therefore rely on other methods as well.}.   But the boundary condition for (quasi-)normal modes of scalar-sector master fields turn out to explicitly depend on the frequency.  This leads to a very different type of eigenvalue problem than is familiar from the study of spin-0 or spin-1 fields and can lead to instabilities even when both tensor- and vector-modes are completely stable.   As will be described below, the fact that such novel boundary conditions first arise for spin-2 appears to be linked to the second-order nature of the equation of motion.

This article is organized as follows. Section \ref{sec:MF formalism} reviews the master field formalism
of \cite{Kodama2003} and determines the approprate boundary conditions on master fields
at a Dirichlet wall.  We also present an algorithm to reconstruct the full metric perturbation from the master field and its derivatives.  Finally, we show that adding a black hole to our system leads to the existence of new modes that can be obtained by acting with  diffeomorphims but which are nevertheless physical.  These modes are manifestly stable except in those de Sitter cases where the gravitational potential is outward-decreasing at the Dirichlet wall.  Such cases occur when the wall is (sufficiently) closer to the cosmological horizon than to that of the black hole.
Section \ref{sec:S2 results} presents our results for spherical domain walls in Minkowski space, AdS, and dS.
In section \ref{sec: flat wall results} we consider perturbations in the presence of flat walls in pure Minkowski space and in AdS.
We summarize our results in section \ref{sec:discussion}, discussing some of their implications and the relation to certain non-perturbative issues.  A few aspects or our treatment, and in particular analyses of non-diffeomorphism modes in black hole spacetimes,  are relegated to the appendices.

\section{Master field formalism}
\label{sec:MF formalism}

The linear stability of gravitational solutions has long been an important subject of investigation.   Early studies date back to \cite{Regge:1957td,Zerilli:1971wd,Zerilli:1970se}, where four-dimensional gravitational perturbations where decoupled. With the aid of these techniques, the stability of Schwarzschild black holes under linear gravitational
fluctuations was finally established in \cite{Vishveshwara:1970cc,Price:1972pw,1979JMP....20.1056W}.
Over the years, similar formalisms to explicitly analyze linear stability have been developed
\cite{Bardeen:1972fi,Teukolsky:1972my,Cardoso:2001bb,Cardoso:2001vs}, all based on reducing
the system of perturbations to decoupled scalar equations.

The most general such formalisms for both neutral and electrically charged spacetimes are due to Kodama and Ishibashi \cite{Kodama2003}, \cite{Kodama:2003kk}. These allow one to
deal in a gauge invariant way with perturbations in arbitrary dimensions, in spacetimes that are warped products over any maximally-symmetric space, for all values of the cosmological constant.
Making use of this decoupling, many black hole spacetimes with different asymptotics have been shown to be linearly stable
\cite{Kodama:2003kk,Ishibashi:2003ap,Kodama:2003ck,Kodama:2004gz}.  Alternatively, it is sometimes possible to use local Penrose inequalities to argue that some instability must exist \cite{GibbonsPhD,Figueras:2011he,Hollands:2012sf}  though we will not use this method below.

This section reviews the master field technology of \cite{Kodama2003} and expresses our Dirichlet boundary conditions on metric perturbations
in this language.  We also reconstruct the metric fluctuations in terms of the master fields, making it explicit that
for a given solution of the master field equation satisfying the appropriate boundary condition there is a linearized solution of Einstein's equations satisfying our Dirichlet boundary condition at the wall.  Finally, we study the pure-diffeomorphism modes that arise when places a black hole inside a spherical cavity formed by a Dirichlet wall.

\subsection{Review}

Consider an $(n+2)$-dimensional metric of the form
\begin{equation}\label{general g}
	ds^2 = \hat g_{\mu \nu} dx^\mu dx^\nu =  g_{ab}(y) dy^a dy^b + r^2(y) d \sigma_n^2,
\end{equation}
\noindent where $g_{ab}$ is a two-dimensional Lorentzian metric and $d \sigma_n^2 = \sigma_{ij} dz^i dz^j$ is the metric
on the maximally symmetric base space with unit curvature $K = 0, +1$. We leave the case $K=-1$ for
future work\footnote{Black holes with hyperbolic horizons have previously been studied in \cite{Gibbons:2002pq}, and their
linear stability was recently established in \cite{Birmingham:2007yv}.}. Coordinates  $x^\mu$ labelled by
Greek indices are space-time coordinates, which we split them into base coordinates, $z^i$, and orbit coordinates $y^a$.
We will be specifically interested in static solutions of vacuum gravity with metric
\begin{equation}\label{Schw background}
	ds^2 = - f(r) dt^2 + \frac{dr^2}{f(r)} + r^2 d \sigma_n^2, \qquad f(r) = K - \frac{2 M}{r^{n-1}} - \lambda r^2.
\end{equation}
Here $\sigma_{ij}$ is the metric on a $n$-dimensional unit sphere or an $n$-dimensional plane. In the former case,
Einstein's equations fix $K= 1$, while in the latter they imply $K= 0$. The constants $M$ and $\lambda$ are related to the total
energy $E$ and the cosmological constant $\Lambda$ by
\begin{equation}
	E = \frac{n M {\cal A}_n}{8 \pi G}, \qquad \Lambda = \frac{n(n+1) \lambda}{2}, 	
\end{equation}
\noindent where
\begin{equation}
 	{\cal A}_n = \int d^n y \sqrt{\sigma}
\end{equation}
\noindent is the area of the base space. The overall strategy of \cite{Kodama2003} is to organize the perturbations of
\eqref{Schw background} according to how they transform under isometries of $\sigma$. In practice, this amounts
to writing a generic perturbation as
\begin{equation}
 	h_{\mu \nu} = h^{(S)}_{\mu \nu} + h^{(V)}_{\mu \nu} + h^{(T)}_{\mu \nu},
\end{equation}
\noindent where $S$, $V$ and $T$ indicate the scalar, vector and tensor parts of the metric, each being a sum of
terms of the form
\begin{align}\label{general perturbation}
\nonumber
	h_{ab}^{(S)} &= f_{ab}^{(S)} \ms, \qquad &h_{ai}^{(S)} & = r f_a^{(S)} \ms_i, \qquad & h_{ij}^{(S)} &= 2 r^2 ( H_L^{(S)} \sigma_{ij} \ms + H_T^{(S)} \ms_{ij}  ),
\\
	h_{ab}^{(V)} &= 0, \qquad & h_{ai}^{(V)}& = r f_a^{(V)} \mv_i, \qquad & h_{ij}^{(V)} &= 2 r^2  H_T^{(V)} \mv_{ij},
\\
\nonumber
	h_{ab}^{(T)} &= 0, \qquad & h_{ai}^{(T)}& = 0, \qquad & h_{ij}^{(V)} &= 2 r^2  H_T^{(T)} \mt_{ij},
\end{align}
\noindent where $\ms$, $\mv_i$,  $\mt_{ij}$ denote scalar, vector and tensor harmonics, which are functions
on $\sigma$ that satisfy
\begin{align}
	({\cal D}^2 + k_S^2) \ms & = 0, \qquad   \\
	({\cal D}^2 + k_V^2) \mv_i &= 0, \qquad &{\cal D}_i \mv^i& = 0\\
	({\cal D}^2 + k_T^2) \mt_{ij} &= 0, \qquad &{\cal D}_i \mt^{j}& \hs _i = 0, \qquad &\mt^i \hs _i& = 0.
\end{align}
Here ${\cal D}_i$ is the covariant derivative associated to $\sigma$ with ${\cal D}^2 = {\cal D}_i {\cal D}^i$ the corresponding Laplacian.
The eigenvalues $k_S^2$, $k_V^2$ and $k_T^2$ are non-negative. For $K = 0$ they form a continuous spectrum,
while for $K=1$ they are discrete and take the form
\begin{align}
 	k_S^2 &= \ell_S (\ell_S + n - 1), \qquad &\ell_S& = 0,1,2 \ldots   \\
 	k_V^2 &= \ell_V (\ell_V + n - 1 ) - 1, \qquad &\ell_V& = 1,2, \ldots \\
 	k_T^2 &= \ell_T (\ell_T + n - 1 ) - 2, \qquad &\ell_T& = 1,2, \ldots
 \end{align}

As noted in \cite{Kodama2003}, the modes for which $k_S^2 = 0$ and $k_S^2 = n K$ are special and we will discuss them separately below.
We shall refer to modes with $k_S^2 (k_S^2 - n K) \neq 0$ as {\it generic} modes. For these perturbations, the
remaining quantities $\ms_i$, $\ms_{ij}$, $\mv_{ij}$ appearing in \eqref{general perturbation} can be written in terms of the basic harmonics
as
\begin{equation}
	\ms_i = - \frac{1}{k_S} {\cal D}_i \ms, \qquad \ms_{ij} = \frac{1}{k_S^2} {\cal D}_i {\cal D}_j \ms + \frac{1}{n} \sigma_{ij} \ms,
	\qquad \mv_{ij} = - \frac{1}{k_V} {\cal D}_{(i} \mv_{j)}.
\end{equation}
The symmetries imply that perturbations belonging to different sectors decouple from each other and can be studied independently.
To understand the effect of gauge transformations, note that infinitesimal diffeomorphisms generated by a vector field $\xi_\mu$ may be similarly decomposed into sums of scalar and vector pieces $\xi_\mu^{(S)}$, $\xi_\mu^{(V)}$ of the form
\begin{align}\label{diffeo split}
	\xi_a^{(S)} &= c_a^{(S)}(t,r) \ms, \qquad &\xi_i^{(S)}& =  L^{(S)}(r,t) r \ms_i \\
	\xi_a^{(V)} &= 0, \qquad &\xi_i^{(V)}&  = L^{(V)}(r,t) r \mv_i.
\end{align}
There are no tensor diffeomorphisms so $H_T^{(T)}$ is diffeomorphism invariant. To extract diffeomorphism-invariant information
from the scalar and vector sectors, it is useful to introduce the quantities
\begin{align}\label{gauge inv Fs}
	F^{(S)} &= H_L^{(S)} + \frac{1}{n} H_T^{(S)} + \frac{1}{r} (D^a) G_a, \qquad F_{ab}^{(S)} = f_{ab}^{(S)} + 2 D_{(a} G_{b)} \\
	F_a^{(V)} &= f_a^{(V)} + \frac{r}{k_V} D_a H_T^{(V)},
\end{align}
\noindent where
\begin{equation}
	G_a = \frac{r}{k_S} \left( f_a^{(S)} + \frac{r}{k_S} D_a H_T^{(S)} \right),
\end{equation}
\noindent and $D_a$ is the covariant derivative compatible with the metric $g_{ab}$ introduced in \eqref{general g}.
One may then show \cite{Kodama2003} that \eqref{gauge inv Fs} are left invariant by diffeomorphisms generated by vectors
of the form \eqref{diffeo split}.

The main result of \cite{Kodama2003} is that the dynamics of the generic modes of each sector is encoded in a single scalar equation
of the form
\begin{equation}
 	\Box_g \Phi^{(I)}(r,t) =  V_I(r) \Phi^{(I)}(r,t),
\end{equation}
for gauge-invariant so-called master fields $\Phi^{(I)}$, where $\Box_g$ is the D'Alambertian for the metric $g_{ab}$ and the index $I$ denotes the scalar, vector or tensor sector.
The potential terms $V_I(r)$ in the master field equation are different for each sector.
The master fields are given in terms of the basic gauge invariants defined above by
\begin{align}
\label{Phi S}
	\Phi^{(S)} &= \frac{2 n r^{n-1} F^{(S)} - n r^{n-2} \partial_t F^{(S) r} \hs_t}{r^{n/2-1} H(r)}, \\
\label{Phi V}
	\epsilon^{ab} D_b (r^{n/2} \Phi^{(V)}) &= r^{n-1} F^{(V) a}, \\
\label{Phi T}
	\Phi^{(T)} &= r^{n/2} H_T^{(T)},
\end{align}
\noindent where
\begin{equation}
	H(r) = m  + \frac{n(n+1)}{2} x,
\end{equation}
\noindent with
\begin{equation}
	m = k_S^2 - n K, \qquad x = \frac{2 M}{r^{n-1}}.
\end{equation}
The potential $V_I$ for each sector is
\begin{align}\label{VS}
\nonumber
	V_S(r) &= \frac{1}{16 r^2 H^2} \{  - [ n^3(n+2)(n+1)^2 x^2 - 12 n^2 (n+1)(n-2) m x + 4 (n-2)(n-4) m^2  ]y  \\
\nonumber
				&+ n^4 (n+1)^2 x^3 + n(n+1)[  4 (2 n^2 - 3 n + 4) m + n (n-2)(n-4)(n+1) K  ] x^2 \\
				&- 12 n [  (n-4)m + n (n+1) (n-2) K ] m x + 16 m^3 + 4 K n (n+2) m^2 \}, \\
\label{VV}
	V_V(r) &= \frac{1}{r^2} \left[  K + k_V^2 + \frac{n(n-2) K}{4} - \frac{n(n-2)}{4} \lambda r^2 - \frac{3n^2 M}{2 r^{n-1}}  \right], \\
\label{VT}
	V_T(r) &= \frac{1}{r^2} \left[  2 K + k_T^2 + \frac{n(n-2) K}{4} - \frac{n(n+2)}{4} \lambda r^2 + \frac{n^2 M}{2 r^{n-1}}  \right],
\end{align}
\noindent where $y = \lambda r^2$.  Due to the lack of diffeomorphism-invariant gravitational fluctuations in low dimensions, the tensor master fields vanish identically for $n \le 2$ while the vector and scalar master fields vanish identically for $n \le 1$.

Fourier expanding in the time direction with $\Phi(r,t) = e^{- i \omega t} \phi(r)$, we obtain
\begin{equation}\label{KG master field}
	\partial_r( f \partial_r \phi^{(I)}) + f^{-1} \omega^2 \phi^{(I)} = V_{I} \phi^{(I)}.
\end{equation}
For generic modes, equation \eqref{KG master field} will be our starting point to study stability.

We now return to the special scalar modes $k_S^2 (k_S^2 - n K) = 0$.
For $k_S = 0$, $\ms$ is constant, and $\ms_i$, $\ms_{ij}$ are not defined. This
mode simply shifts the values of $M$ and $\lambda$, so it need not be considered further. For $K=1$ the mode $k_S^2 = n K$ is clearly distinct from the mode $k_S = 0$, and it turns out to have $\ell_S = 1$.
As noted in \cite{Kodama2003}, these modes are generated by diffeomorphisms. But they may still contain physical information due to the presence of the wall\footnote{These modes are trivial (pure gauge) in the absence of the wall.  Perhaps for this reason, they appear not to have been previously studied in detail.}. We will postpone the analysis of this type of excitation until section \ref{sec:Scalar modes}.

\subsection{Boundary conditions in terms of Master Fields}

As we now demonstrate,
the problem of studying the linearized Einstein equation in the presence of a Dirichlet wall is equivalent to solving \eqref{KG master field} with
particular boundary conditions. Examining \eqref{general perturbation}, we see that fixing the induced metric on the surface $r = r_D$ is equivalent to requiring
\begin{equation}\label{metric D bcs}
f^{(S)}_{tt}  =  f^{(S)}_{t} = H_T^{(S)} =  H_L^{(S)} = 0, \quad f^{(V)}_{t}  = H_T^{(V)} = 0, \quad H_T^{(T)} = 0,
\end{equation}
\noindent at $r = r_D$. Note that $h_{\mu r}$ remains unrestricted by our boundary conditions.

We now derive the appropriate boundary conditions on the master fields which implement \eqref{metric D bcs} in each sector. We also describe how to recover a metric perturbation that vanishes at the Dirichlet wall from the resulting master fields.

\subsubsection{Tensor modes}

For the tensor modes, we see from \eqref{Phi T} that the master field boundary condition is simply
\begin{equation}\label{bc MF tensor}
	\phi^{(T)}(r_D) = 0.
\end{equation}
\noindent Given a solution of the master field equation, the tensor metric perturbation is just
$H_T^{(T)} = r^{-n/2} \Phi^{(T)}$ with all other fields in \eqref{metric D bcs} set to zero.
%

\subsubsection{Vector modes}

For vector modes, one may note that the boundary condition \eqref{metric D bcs} implies
\begin{equation}
	F_t^{(V)}(r_D) = f_t^{(V)} - i \omega \frac{r H_T^{(V)}}{k_V} \bigg|_{r = r_D} = 0.
\end{equation}
Equation \eqref{Phi V} then requires,
\begin{equation}\label{bc MF vector}
	\partial_r(r^{n/2} \phi^{(V)}) |_{r = r_D} = 0,
\end{equation}
\noindent which is the desired boundary condition on vector master fields.

To reconstruct a metric perturbation we use the single gauge parameter available in
the vector sector \eqref{diffeo split} to set
\begin{equation}
	f_t^{(V)} = 0.
\end{equation}
We then define
\begin{align}
	H_T^{(V)} &= - \frac{k_V}{i \omega r} F_t^{(V)} \\
	f_r^{(V)} &= F_r^{(V)} - \frac{r}{k_V} \partial_r H_T^{(V)},
\end{align}
\noindent where the $F_a^{(V)}$ are meant to be understood as functions of $r$, $\phi^{(V)}$, and $\partial_r \phi^{(V)}$,
as specified by \eqref{Phi V}. The resulting metric perturbation solves the linearized Einstein equation and satisfies our boundary condition.

\subsubsection{Generic scalar modes}
\label{sec:Scalar modes}

We finally turn to the generic scalar modes. These include all scalar modes with $K=0$ and the $K=1$
scalar modes for which $\ell_S > 1$.
Consider the gauge invariant quantities $X$, $Y$, $Z$ given implicitly by
\begin{align}\label{def XYZ}
\nonumber
	F^{(S) t} \hs _t &= \frac{(n-1) X - Y}{n r^{n-2}}, \qquad  &F^{(S) r} \hs _r& = \frac{(n-1) Y - X}{n r^{n-2}}, \\
	F^{(S) r} \hs _t &= \frac{Z}{r^{n-2}}, \qquad &F^{(S)}& = - \frac{X+Y}{2 n r^{n-2}}.
\end{align}
Note that they satisfy the constraint
\begin{equation}\label{Faa and F constraint}
	F^{(S) a} \hs _a + 2 (n-2) F^{(S)} = 0,
\end{equation}
\noindent which allows us to invert the relations \eqref{def XYZ} to algebraically express
$X$, $Y$ and $Z$ in terms of $F_{ab}^{(S)}$ and $F^{(S)}$. It follows from \eqref{Phi S} and
\eqref{def XYZ} that
\begin{equation}\label{MF scalar def}
	\phi^{(S)} =  \frac{i n \omega Z - r (X+Y)}{r^{n/2-1} H}.
\end{equation}
Moreover, as shown in \cite{Kodama2003}, using Einstein's equations we can express $X$, $Y$ and $Z$
in terms of $\phi^{(S)}$ and $\partial_r \phi^{(S)}$. We refer the reader to \cite{Kodama2003} for the explicit expressions.

It will prove convenient to introduce
\begin{equation}
	E^{\pm} := F^{(S) t} \hs _t \pm \frac{r f'}{f} F^{(S)}.
\end{equation}
A simple calculation then gives
\begin{equation}
	E^- = - \frac{1}{f} f_{tt}^{(S)} + \frac{2 i \omega r}{k_S f} f^{(S)}_t +
	\frac{1}{f} \left[  2 \left( \frac{\omega r}{k_s} \right)^2 - \frac{r f'}{2}  \right] H_T^{(S)} - \frac{r f'}{f} H_L^{(S)}.
\end{equation}
We now see from \eqref{metric D bcs} that the Dirichlet boundary condition requires $E^{-} = 0$ at $r = r_D$. From \eqref{def XYZ},
this means that the scalar master field must satisfy the boundary condition
\begin{equation}\label{Dbc Eminus}
 	E^-(r_D) = \frac{1}{2} \left(  (n-1)  + \frac{r f'}{2 f} \right) X - \frac{1}{2} \left(  1  + \frac{r f'}{2 f} \right) Y	\bigg|_{r = r_D} = 0,
\end{equation}
Recall that $X$ and $Y$ can be written in terms of $r$, $\phi^{(S)}$ and $\partial_r \phi^{(S)}$. Hence, the boundary condition \eqref{Dbc Eminus} implies a Robin boundary condition for $\phi^{(S)}$ of the form
\begin{equation}\label{bc MF scalar}
	\alpha(r) \partial_r \phi^{(S)} + \beta(r) \phi^{(S)} |_{r = r_D} = 0
\end{equation}
\noindent where
\begin{align}
	\alpha &= -\frac{m (n-1) f^2}{r H} \\
	\beta &= \omega^2 + \frac{k_S^2(k_S^2 - n f - H)}{n^2 r^2} - \frac{m(n-1) (2 k_S^2 + n (n-2) f ) f }{2 n r^2 H}	- \frac{m (n-1) H' f^2}{r H^2}.
\end{align}
Note that the first term in $\beta$ depends explicitly on the frequency $\omega$.  This novel feature appears to arise from the fact that the dictionary relating scalar perturbations to the (tensor) metric field contains extra derivatives, and from the fact that two radial derivatives are related to second time derivatives by the equation of motion.

It is not yet manifest that the boundary condition \eqref{Dbc Eminus} is truly equivalent to fixing the induced metric at $r = r_D$. We now show that
this is so by giving an algorithm to construct a scalar metric perturbation $h_{\mu \nu}^{(S)}$ which
solves the linearized Einstein's equations and satisfies our Dirichlet boundary condition.
First we must choose a gauge. A convenient choice is to set
\begin{equation}\label{gauge choice}
	f_t^{(S)} = H_T^{(S)} = H_L^{(S)} = 0.
\end{equation}
This exhausts the three functions in the general scalar gauge transformation in \eqref{diffeo split}. Now let the remaining
metric functions be given by
\begin{align}\label{scalar hmunu ito gauge inv}
\nonumber
	f_{tt}^{(S)} &= - f E^-, \\
\nonumber
	f_r^{(S)} &= \frac{k_S}{2f} F^{(S)} , \\	
\nonumber
	f_{tr}^{(S)} &= F_{tr}^{(S)} + \frac{i \omega r}{k_S} f_r^{(S)}, \\
	f_{rr}^{(S)} &= F_{rr}^{(S)} - 2 D_r \left(  \frac{r f_r^{(S)}}{k_S} \right),
\end{align}
\noindent where in these expressions $E^{\pm}$, $F_{tr}^{(S)}$, $F_{rr}^{(S)}$ are meant to be taken as functions of $X$, $Y$ and $Z$
according to \eqref{def XYZ}. Given a solution of the master field equation \eqref{KG master field} for the scalar potential, we can calculate
$X$, $Y$, $Z$ using the explicit expressions in \cite{Kodama2003}. We then define our metric functions by \eqref{gauge choice}
and \eqref{scalar hmunu ito gauge inv}. The resulting metric is a solution to the linearized Einstein's equations which satisfies our Dirichlet boundary condition when the master field satisfies \eqref{bc MF scalar}.

\subsection{$\ell_S = 1$ modes}
\label{sec:l=1 modes}

As mentioned above, modes with $K=1$, $\ell_S = 1$ must be considered separately since they yield $\ms_{ij} = 0$ identically, so that the function $H_T^{(S)}$ drops out of the analysis above.
As noted in \cite{Kodama2003}, all solutions with $\ell_S = 1$ can be generated by diffeomorphims.  Here we may allow any $n \ge 0$.
A general $\ell_S = 1$ diffeomorphim of the form \eqref{diffeo split} generates the perturbation with $\ms_{ij} = 0$
and, dropping the $(S)$ superscripts for the rest of this section since we consider only scalar modes, takes the form
\begin{align}
	f_r &= \frac{r L' - L - \sqrt{n} c_r}{r},\\
	f_{rr} &= \frac{c_r f' + 2 f c_r'}{f}, \\
	f_{rt} &=  \frac{f c_t' - c_t f' - i \omega f c_r}{f}, \\
	f_{tt} &= - ( c_r f f' + 2 i \omega c_t ),  \\
	f_t &= - \frac{i \omega r L + \sqrt{n} c_t}{r}, \\
	H_L &= \frac{\sqrt{n} L + n f c_r}{n r}. \\
\end{align}
Thus for $\ell_S = 1$, the Dirichlet boundary condition requires
\begin{align}
	i \omega r_D L(r_D) + \sqrt{n} c_t(r_D) &=0, \\
	\sqrt{n} L(r_D) + n f(r_D) c_r(r_D) &= 0, \\
\label{Dbc w2 Phi=0}
	(\omega^2 - \omega_{\ell_S = 1}^2) L(r_D) &= 0, 	
\end{align}
\noindent where
\begin{equation}\label{w2 l=1}
	\omega_{\ell_S = 1}^2 = \frac{f'(r_D)}{2 r_D}.
\end{equation}
Consider first the case in which we satisfy \eqref{Dbc w2 Phi=0} by setting
$L(r_D) = 0$. The mode then corresponds to a diffeomorphism that vanishes at $r=r_D$.
We expect this perturbation to be pure gauge since acts trivially on the boundary
(see e.g. \cite{Fischetti:2012rd}, or \cite{Brown:1992br} for a well-known analogous case).
In appendix \ref{app:symplectic}, we will see by explicit computation
in four dimensions that diffeos with $L(r_D) = 0$ are indeed null directions of the symplectic structure and
can thus be regarded as pure gauge. We emphasize that our analysis both above and below is independent of the sign of the wall's extrinsic curvature and thus applies equally well on either the ``inside'' or the ``outside'' of our wall.

On the other hand, one may instead solve \eqref{Dbc w2 Phi=0} by imposing $\omega = \pm \omega_{\ell_S =1}$.
We can of course also satisfy any linear boundary condition at the other end of spacetime by choosing the diffeomorphism to vanish in that vicinity.  The result is a linearized solution generated by a diffeomorphism that acts non-trivially on the wall. Such modes can be physical, and for $M \neq 0$ one may verify that this is indeed the case by showing that they have non-zero symplectic inner products. Since it differs qualitatively from the rest of this work, we have relegated this calculation to appendix \ref{app:symplectic}.  There it is performed explicitly for the four-dimensional case ($n=2$), but the result should be similar in other dimensions.   In contrast, even modes with $\omega = \pm \omega_{\ell_S =1}$ are pure gauge for $M= 0$.  So for $M \neq 0$ they are naturally interpreted as describing the motion of the black hole's center of mass relative to the spherical wall.  Note that they are stable when the gravitational potential $f$ increase at the wall in the same direction as the radius $r$, while they are unstable when $f$ decreases in this direction.

For $M\neq 0$, $\lambda \le 0$ it is straightforward to show that \eqref{w2 l=1} is always positive.  So in such cases the black hole's center of mass executes stable oscillations.  This confirms expectations from \cite{Andrade2015moduli}, which argued that uncharged black holes are repelled from Dirichlet walls.

But \eqref{w2 l=1} can become negative for $\lambda > 0$ where we find
\begin{align}\label{w inst Sch dS}
\nonumber
	\omega^2 &= \frac{f'(r_D)}{2 r_D}  \\
		&= \frac{1}{2 \rho_D \rho_H L_{dS}^2} \left[ (n-1) \left( \frac{\rho_H}{\rho_D} \right)^n (1 - \rho_H^2) - 2 \rho_D \rho_H \right]
\end{align}
for $\rho_D = r_D/L_{dS}$ and $\rho_H = r_H/L_{dS}$, and where we require $\rho_H < \rho_D < 1$ so that the wall is timelike (and lies inside the de Sitter horizon).  The resulting instability is no surprise for $\rho_H \ll \rho_D$, where it is clear that the black hole experiences the same instability as geodesics in de Sitter which can slide off the central maximum of the effective potential toward the de Sitter horizon.   And it is reasonable that this instability will become stronger as the edge of the black hole approaches the de Sitter horizon (when $\rho_H$, and thus also $\rho_D$, approaches $1$).
Noting that \eqref{w inst Sch dS} is positive for $\rho_D = \rho_H$ shows that bringing the wall close enough to the black hole always removes the instability.  This again verifies the repulsive interaction between the wall and the black hole horizon.

%

\section{Spherical walls}
\label{sec:S2 results}

We will first address the case of a spherical wall at $r = r_D$ in Minkowski, AdS, and dS.  For each case, we study  separately the fluctuations ``inside'' and ``outside'' the Dirichlet wall, i.e. we consider the
spacetimes for which the metric is \eqref{Schw background} with $M=0$ and $K=1$, restricted to the region $r < r_D$ and then to $r > r_D$. In the de Sitter case, we consider only the static region inside the de Sitter horizon.   Since the modes generated by diffeomorphisms were analyzed in section \ref{sec:l=1 modes} and appendix \ref{app:symplectic}, we confine ourselves here to modes with $\ell_S > 1$.

In addition to the boundary condition at $r_D$, we impose regularity at $r=0$, outgoing boundary conditions at Minkowski  $r=\infty$ and the de Sitter horizon, and asymptotically AdS boundary conditions at AdS $r=\infty$.  Thus both the Minkowski and dS cases define quasi-normal mode problems rather than normal mode problems.  The asymptotically AdS boundary condition fixes the the conformal metric at $r=\infty$; for gravitational fluctuations, this condition can be obtained by carefully taking the $r\rightarrow \infty$ limit of the same boundary condition that
we impose at our Dirichlet wall.  Other boundary conditions appear to lead to ghosts \cite{Andrade:2011dg,Compere:2008us}.

\subsection{Minkowski space}
\label{sec:results mink}

We begin with Minkowski space ($K=1$,
$\lambda = 0$, $M=0$ in \eqref{Schw background}). For this background, all three master field potentials are positive definite and can be written
\begin{equation}
 	V_I = \frac{1}{4 r^2} [ 4 \ell_I (\ell_I + n-1) + n (n-2) ].
\end{equation}
The wave equation can be solved analytically, but for general modes inside the wall -- and for tensor modes outside -- one may also show stability by extending the argument of \cite{Ishibashi:2003ap} and keeping track of the new
boundary terms at $r_D$ that arise when integrating by parts.

\subsubsection{Inside the wall}
\label{MinkInside}

Consider first the region $0 \le r< r_D$. In this case we begin
by introducing the radial coordinate $r_* = \int f^{-1}dr$ in \eqref{KG master field}, finding
\begin{equation}\label{proof M=0 1}
 	- \frac{d^2}{d r_*^2} \phi^{(I)} + f V_I \phi^{(I)} =  \omega^2 \phi^{(I)} 	.
\end{equation}
Recall that our boundary conditions are regularity at the origin and, depending on the sector, we have \eqref{bc MF tensor}, \eqref{bc MF vector},
or \eqref{bc MF scalar} at the Dirichlet wall.
Studying the wave equation explicitly shows that the two characteristic behaviours at the origin are
$\phi^{(I)} \sim r^{n/2+ \ell_I}$ and $\phi^{(I)} \sim r^{1 - n/2 - \ell_I}$.  Thus regularity at $r = 0$ implies
$\phi^{(I)}(r=0) = 0$.
Following \cite{Ishibashi:2003ap} we multiply \eqref{proof M=0 1} by the complex conjugate of $\phi^{(I)}$, which we denote by $\bar \phi^{(I)}$,
and integrate both sides. After integrating by parts we obtain
\begin{equation}\label{mink LSP}
	\int_{r_*(0)}^{r_*(r_D)} ( |\partial_{r_*} \phi^{(I)}|^2 + f V_I |\phi^{(I)}|^2  ) dr_* - \bar \phi^{(I)} f \partial_r \phi^{(I)} \bigg|_{0}^{r_D} =
	\omega^2 \int_{r_*(0)}^{r_*(r_D)} |\phi^{(I)}|^2 dr_* .
\end{equation}
The boundary conditions in all three sectors can be written
\begin{equation}\label{gen bc at wall}
	[ a_I(r, k_I) \partial_r \phi^{(I)}	+ b_I(r, k_I) \phi^{(I)} - \omega^2 c(r, k_I) \phi^{(I)} ] \big |_{r = r_D} = 0,
\end{equation}
\noindent where $a_I(r, k_I)$, $b_I(r, k_I)$ and $c_I(r, k_I)$ are functions of the radial coordinate and the radial coordinate and the
spatial momentum.
For the tensor modes, $b_T = 1$, $a_T = c_T = 0$, so it immediately follows from \eqref{mink LSP}
that $\omega^2 \geq 0$ for any non-vanishing mode.
For the vector and scalar modes,  we can use the boundary condition at the wall \eqref{gen bc at wall} in \eqref{mink LSP} to arrive at
\begin{align}\label{LSP vector scalar}
\nonumber
	\int_{r_*(0)}^{r_*(r_D)} ( |\partial_{r_*} \phi^{(I)}|^2 + f V_S |\phi^{(I)}|^2  ) dr_* +
	 \frac{b_I(r_D,k_I)}{a_I(r_D,k_I)} f(r_D) |\phi^{(I)}(r_D)|^2  = \\
	\omega^2 \left( \int_{r_*(0)}^{r_*(r_D)} |\phi^{(I)}|^2 dr_*  + \frac{c_I(r_D,k_I)}{a_I(r_D,k_I)}  f(r_D) |\phi^{(I)}(r_D)|^2 \right)
\end{align}
For the vectors we have
\begin{equation}
	a_V = r, \qquad b_V = \frac{n}{2}, \qquad c_V = 0,
\end{equation}
\noindent so using \eqref{LSP vector scalar} we can conclude that these modes are all stable. Finally, for the scalars
it turns out that
\begin{equation}
	a_S = \frac{(n-1)}{r}, \qquad b_S = \frac{(n-1)}{2 n r^2} [2 k_S^2 + n (n-2)], \qquad c_S = 1,
\end{equation}
\noindent which also leads to $\omega^2 > 0$.
Thus pure Minkowski space is stable against all gravitational perturbations that live inside a spherical Dirichlet
cavity of finite radius.

\subsubsection{Outside the wall}
\label{MinkowskiOutside}

Consider now the region $r > r_D$.   We impose outgoing boundary conditions at $r = \infty$, which selects the profiles that
behave as $\phi^{(I)} \sim e^{+i \omega r}$. The above argument can again be used to show stability of the tensor modes since their boundary term vanished at $r=r_D$ and the outgoing boundary condition requires any unstable mode to vanish exponentially as $r \rightarrow \infty$.  But since $r = r_D$ is now the
lower limit of integration, for vector and scalar modes the boundary term in \eqref{mink LSP} now yields contributions with the wrong sign.  For these modes we  analyze the wave equation in detail.  The radial profiles can then be written
\begin{equation}\label{mink profile out}
	\phi^{(I)} = \left( \frac{\pi}{2} \right)^{1/2} (\omega r)^{1/2} H^{(1)}_{\nu_I}(\omega r),
\end{equation}
\noindent where $H^{(1)}_{\nu_I}$ is the Hankel function of the first kind and $\nu_I = (n-1)/2 + \ell_I$. The factor of
$(\pi \omega/2)^{1/2}$ is inserted for convenience. It is useful to note that the Hankel functions reduce to products
of exponentials and polynomials for half-integer $\nu_I$. For example
\begin{align}\label{tensor prof 1}
 	\phi^{(I)} &= i \frac{e^{i \omega r}}{\omega^2 r^2}   ( \omega^2 r^2 + 3 i \omega r  - 3),
 	\qquad &n& =2, \ell_I = 2, \\
\label{tensor prof 2}
 	\phi^{(I)} &= \frac{e^{i \omega r}}{\omega^3 r^3}   ( \omega^3 r^3 + 6 i \omega^2 r^2 - 15  \omega r  - 15 i),
 	\qquad  &n& =4, \ell_I = 2,\\
\label{tensor prof 3}
	\phi^{(I)} &= - i \frac{e^{i \omega r}}{\omega^4 r^4}
	( \omega^4 r^4 + 10 i \omega^3 r^3 - 45  \omega^2 r^2  - 105 i \omega r + 105), \qquad &n&=6, \ell_I = 2.
\end{align}
To obtain the desired spectra we must implement the boundary condition at $r=r_D$ numerically. Note that,
because $\phi^{(I)}$ depends on $\omega$ only through the combination $\omega r$, we can define
$\hat \omega = \omega r_D$ and express our solutions in terms of this dimensionless quantity.
In this form the results are independent of $r_D$.

\begin{figure}[h]
\center
\subfigure[][]{
\includegraphics[width=0.4\linewidth]{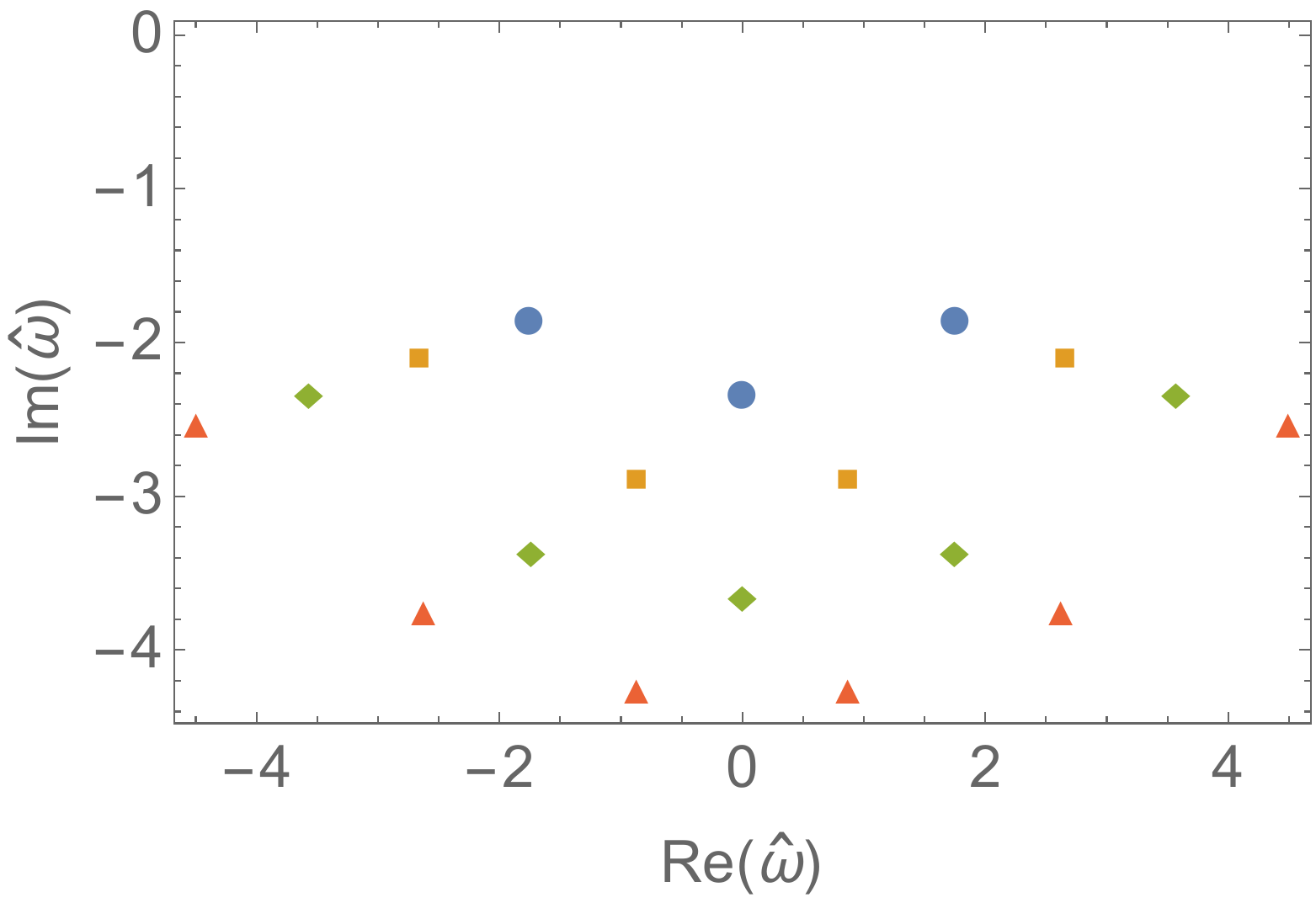}
}\qquad\qquad
\subfigure[][]{
\includegraphics[width=0.40\linewidth]{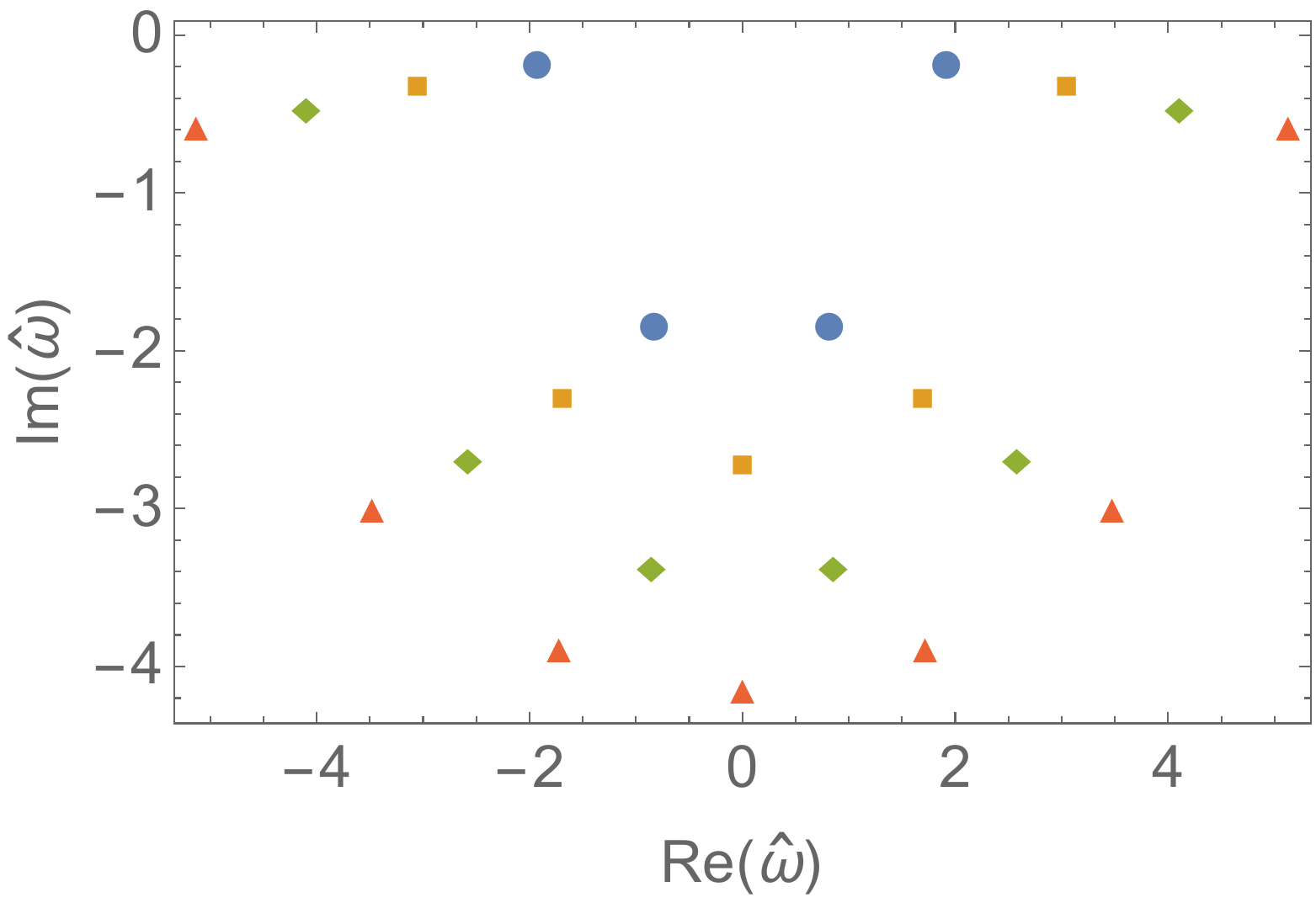}
}
\caption{{\bf External Minkowski tensor and vector modes are stable:} Spectra of frequencies for the tensor (left) and vector (right)
modes for $r> r_D$, $K=1$, $M=\lambda=0$, $n=4$. These are obtained by solving $e_T(\hat \omega)  = 0$
and $e_V(\hat \omega)  = 0$ respectively, with $\ell_I = $ 2, 3, 4, 5 (blue circles, yellow squares, green diamonds, red triangles).}
\label{fig:Mink VyT}
\end{figure}

It is illustrative to begin with the tensor modes, even though these are already known to be stable.  The boundary condition
is simply
\begin{equation}
	e_T(\hat \omega) := \phi^{(T)}(\hat \omega) = 0.
\end{equation}
Solving this numerically in the complex plane, we find that the tensor spectrum consists of a finite number of solutions for each $\nu$ that increases with $\nu$, all of which are stable, i.e. they lie in the lower half plane, see
figure \ref{fig:Mink VyT}.
Note, however, that some frequencies are complex and, from the profiles \eqref{tensor prof 1}-\eqref{tensor prof 3}, this implies that the solutions diverge exponentially at infinity.  This is to be expected since, as remarked above, outgoing boundary conditions at infinity define a quasi-normal mode problem rather than a normal mode problem.

Plugging \eqref{mink profile out} in \eqref{bc MF vector}, we conclude that the vector modes are given by
solutions of
\begin{equation}
\label{eq:eV}
	e_V(\hat \omega) := \hat \omega \phi^{(V)'}(\hat \omega) + \frac{n}{2} \phi^{(V)}(\hat \omega) = 0
\end{equation}
Studying \eqref{eq:eV} in the complex plane one finds that the vector modes
behave similarly to the tensor ones. There are again a finite number of modes for each $\nu$, all of which lie in the lower half plane;
see figure \ref{fig:Mink VyT}.

\begin{figure}[h]
\center
\subfigure[][]{
\includegraphics[width=0.425\linewidth]{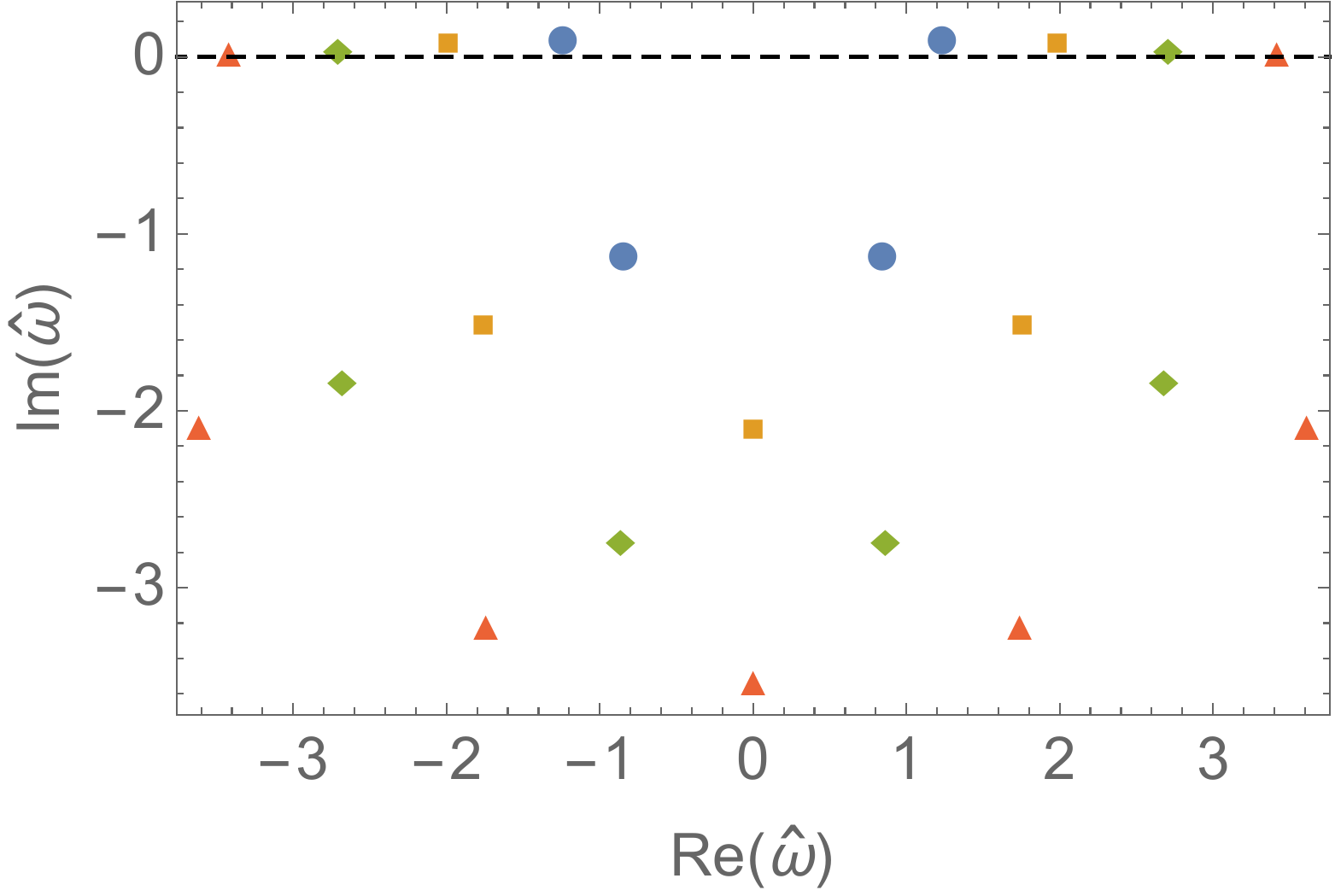}
}\qquad\qquad
\subfigure[][]{
\includegraphics[width=0.44\linewidth]{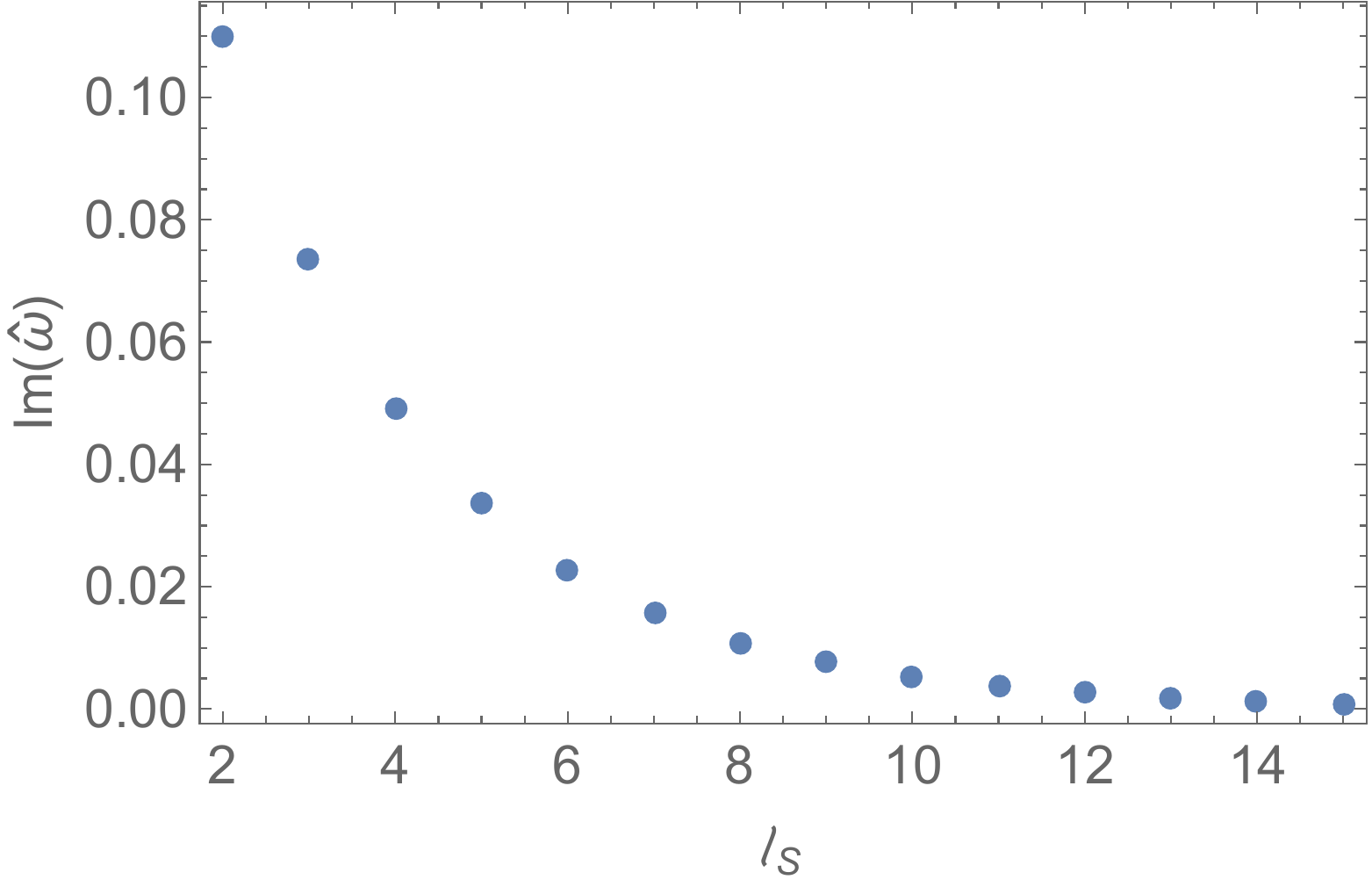}
}
\caption{{\bf External Minkowski scalars are unstable:}  Spectrum of frequencies (left panel) for scalar modes
with $r> r_D$, $K=1$, $M=\lambda=0$, $n=2$. These are given by the solutions of $e_S(\hat \omega)=0$,
with $\ell_S = $ 2, 3, 4, 5 (blue circles, yellow squares, green diamonds, red triangles). On the right panel
we plot the imaginary part of the unstable modes as a function of $\ell_S$.}
\label{fig:Mink S}
\end{figure}

We now turn to the more interesting scalar modes. The boundary condition at the wall \eqref{bc MF scalar}
reduces to
\begin{equation}
\label{eq:eSAdS}
	e_S(\hat \omega):= \hat \omega \phi^{(S)'}(\hat \omega) + \left( \frac{2 k_S^2 + n(n-2)}{2n} -
	\frac{\hat \omega^2}{n-1} \right) \phi^{(S)}(\hat \omega) = 0.
\end{equation}
Here we do find instabilities corresponding to a pair of complex solutions located in the upper half-plane, as illustrated in
figure \ref{fig:Mink S}.

We observe numerically that for any given $(n,\ell_S)$ there
is always a single pair of unstable modes, and that the number of stable modes grows with $\ell$ and $n$.
We have checked this for $n \le 12$ and $2 \le \ell_S \le 15$.
The profiles of the unstable modes decay exponentially
at infinity, so they correspond to normalizable modes which grow exponentially with time.

\subsection{AdS}

We next study Dirichlet walls in AdS ($K=1$, $M=0$,
$\lambda < 0$ in \eqref{Schw background}).   The AdS radius, which we will denote by $L_{AdS}$,
is given by $L_{AdS} = (- \lambda)^{-1/2}$.

\subsubsection{Inside}
The analysis of the interior region $0 \le r < r_D$ proceeds much as in the Minkowski case.  In particular, the linear stability proof of section \ref{MinkInside} applies directly to both tensor and vector modes as the the potentials in \eqref{VT}, \eqref{VV} are again manifestly positive for all $n$ when $\lambda <0$ and their boundary conditions are the same as  in Minkowski space.

The scalars modes must be considered more closely. The potential for the scalar master field reduces to
\begin{equation}
	V_S(r) = \frac{4 k_S^2 + n(n-2)}{4 r^2} + \frac{(n-2)(n-4)}{4 L_{AdS}^2}.
\end{equation}
\noindent For $n\neq 3$ this is positive definite, so in this case we again argue as in \ref{MinkInside}.
The boundary condition \eqref{bc MF scalar} is of the form \eqref{gen bc at wall} with
\begin{align}
\label{aS AdS}
	a_S &= 2 n (n-1) r (1 + r^2/L_{AdS}^2 )^2 \\
\label{bS AdS}
	b_S &=  n (n-2)(n-1)(1 + r^2/L_{AdS}^2 )^2 + 2 k_S^2 [ n(1 + r^2/L^2_{AdS})  - 1]   \\
\label{cS AdS}
	c_S &= 2 n r^2
\end{align}
Because these quantities are all positive, equation \eqref{LSP vector scalar} implies $\omega^2 > 0$ for $n \neq 3$.
It remains to consider the case $n=3$ where $V_S$ is not positive definite.
For general $n$, the solution to the wave equation that is regular at the origin can be written
\begin{equation}
	\phi^{(S)} = \rho^{\ell_S + n/2} (1 + \rho^2)^{\tilde \omega/2} \hs _2 F_1 \left[ \frac{1}{2}(2 + \ell_S + \tilde \omega),
	\frac{1}{2}(n-1 + \ell_S + \tilde \omega); \frac{1}{2} (n+1 + 2 \ell_S ); - \rho^2 \right],
\end{equation}
\noindent with $\rho = r/L_{AdS} $, $\tilde \omega = \omega L_{AdS}$ and $_2 F_1(a,b;c;x)$ a
hypergeometric function.
The frequencies are given by setting $n=3$ and imposing the boundary condition \eqref{gen bc at wall}. For
general $\rho_D = r_D/L_{AdS}$ this needs to be done numerically, but we can obtain an asymptotic expression for
$\rho_D \to \infty$ by expanding the boundary condition near this point, which yields
\begin{equation}
	\alpha(r) \partial_r \phi^{(S)} + \beta(r) \phi^{(S)} |_{r = r_D} =
	\rho^{\ell_S + \tilde \omega - 2}
	\left( \frac{12 \Gamma(\ell_S+2)}{\Gamma[(2+\ell_S+\tilde \omega)/2] \Gamma[(2+\ell_S-\tilde \omega)/2]}  + O (\rho^{-1}) \right)
\end{equation}
Vanishing of the leading term requires
\begin{equation}\label{w AdS asympt}
 	\omega_{\rho_D \to \infty} = \pm ( \ell_S + 2 + 2 p), \qquad p = 0, 1, 2, \ldots
\end{equation}
Now, the potential becomes negative only for $\rho^2 > 4 \ell_S(\ell_S+2) +3 $, which is rather large even for small values of
$\ell_S$.  Combined with stability at $\rho_D = \infty$, this suggests that there should be no instabilities. We have confirmed this by numerically scanning the range
$0.1 \le \rho_D \le 20$, $2 \le \ell_S \le 15$, finding good agreement with \eqref{w AdS asympt}. For example, for $n=2$,
$\rho_D = 6$ we find that the first normal modes are $\pm \tilde \omega = 4.004, 6.032, 8.096, 10.197$.  Thus we conclude AdS inside a spherical cavity to be linearly stable under all gravitational perturbations.

\subsubsection{Outside}

We now consider the region $r \in (r_D, \infty)$. It is convenient to introduce the dimensionless radial coordinate
$ u = \rho^{-1}$, so the range of interest becomes $0 < u < u_D$ with $u_D = 1/\rho_D$.
The fluctuations must preserve the conformal metric at $u=0$, which amounts to imposing
the Dirichlet-wall boundary conditions \eqref{bc MF tensor}, \eqref{bc MF vector}, \eqref{bc MF scalar}
in the limit $u_D \to 0$, see \cite{Dias:2011ss, Michalogiorgakis:2006jc} for an earlier derivation of this condition in the master field formalism.

The profiles that satisfy the boundary condition at $u=0$ take the form
\begin{align}
	\phi^{(T)} &= 	u^{n/2+1} (1+u^2)^{\tilde \omega/2} \hs _2F_1 \left[ \frac{1}{2}(\tilde \omega - \ell_T +2),
	\frac{1}{2} (\ell_T + n+1 + \tilde \omega); \frac{n+3}{2}; - u^2 \right], \\
	\phi^{(V)} &= 	u^{n/2} (1+u^2)^{\tilde \omega/2} \hs _2F_1 \left[ \frac{1}{2}(\tilde \omega - \ell_V +1),
	\frac{1}{2} (\ell_V + n + \tilde \omega); \frac{n+1}{2}; - u^2 \right], \\
	\phi^{(S)} &= 	u^{n/2-1} (1+u^2)^{\tilde \omega/2} \hs _2F_1 \left[ \frac{1}{2}( \tilde \omega- \ell_S ),
	\frac{1}{2} (\ell_S + n-1 + \tilde \omega); \frac{n-1}{2}; - u^2 \right].
\end{align}

The usual plots of where the real and imaginary parts of the $r=r_D$ boundary conditions vanish again show that tensor and vector spectra to consist only of real frequencies (with an infinite set for each $\nu$).  The tensor case constitutes a check on our numerics since, as for $\lambda =0$, we may prove stability as in section \ref{MinkInside}.

\begin{figure}[h]
\begin{center}
\includegraphics[scale=0.5]{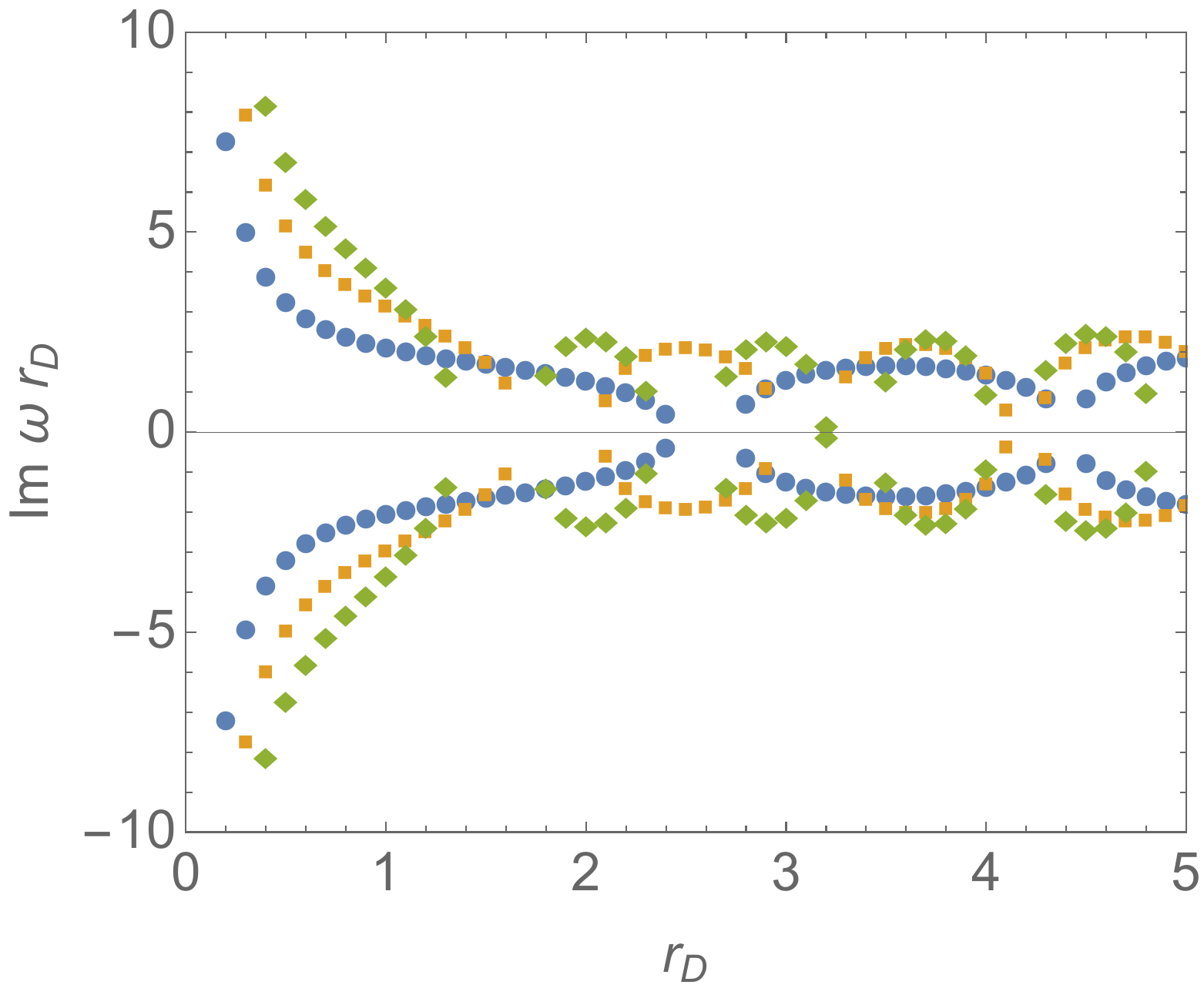}
\caption{The imaginary part of the frequency of the unstable scalar mode outside of a spherical cavity in global AdS$_4$ for
$\ell =2,3,4$ (blue circles, yellow squares, green diamonds).}
\label{fig:AdS islands}
\end{center}
\end{figure}

The behavior in the scalar sector is more complicated.  For a given mode, stability can depend on $u_D$ in an interesting way;
see figure \ref{fig:AdS islands}.  However, for any given $u_D$ it appears that we can find quantum numbers which exhibit instability.
So linearized gravity is unstable outside any spherical Dirichlet wall in global AdS.


\begin{figure}[h]
\begin{center}
\includegraphics[scale=0.5]{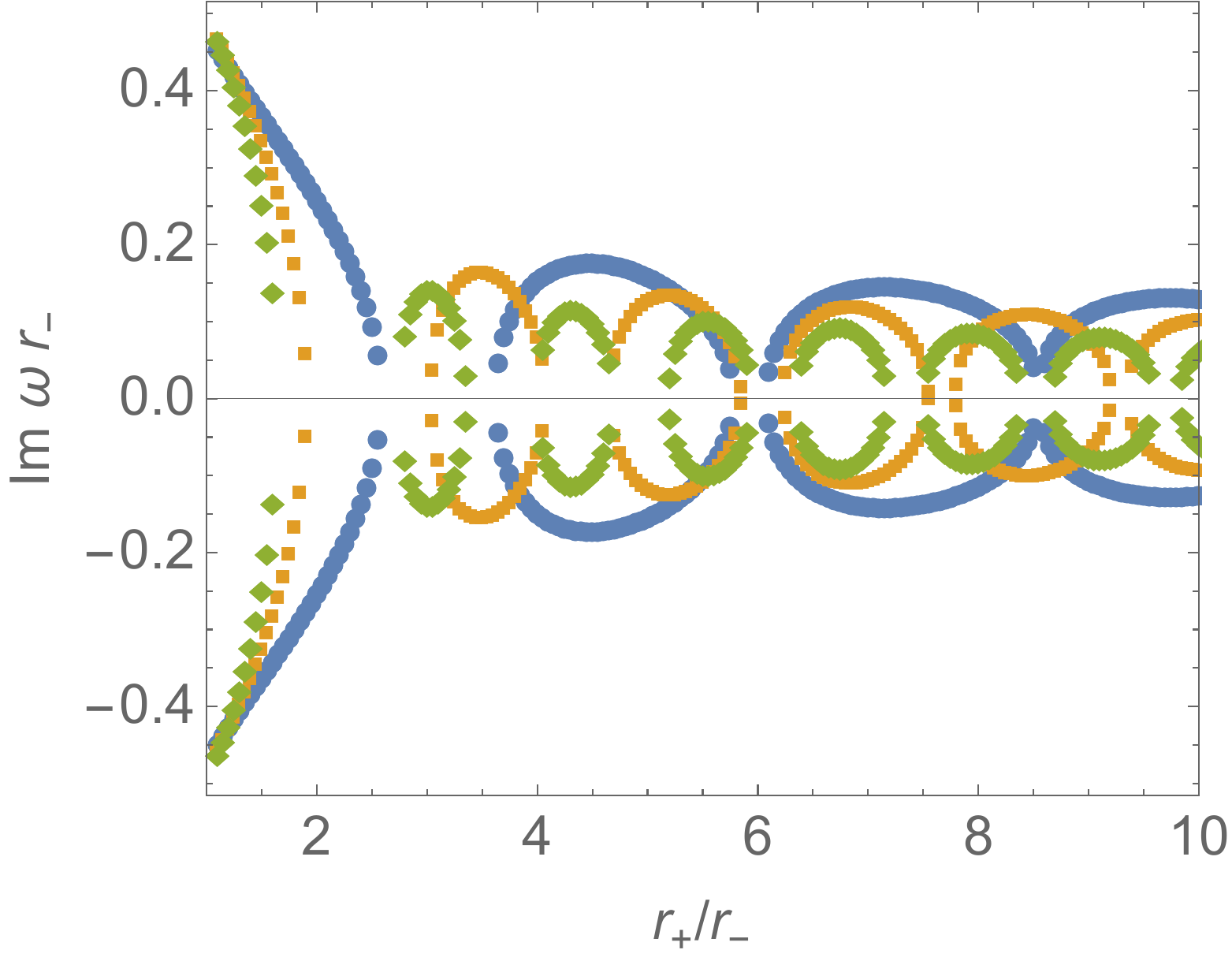}
\caption{ The imaginary part of the frequency of the unstable scalar mode between two cavities in flat
space for $\ell =2,3,4$ (blue circles, yellow squares, green diamonds).}
\label{fig:TwoCavities}
\end{center}
\end{figure}

Since the appropriate boundary condition for an asymptotically AdS spacetime is simply the limit of our Dirichlet boundary condition, it is natural to conjecture that the oscillatory behavior seen between a Dirichlet wall and the AdS boundary would be qualitatively similar to that between two Dirichlet walls.  This is in fact the case and we find similar instabilities in the scalar sector between two spherical walls in Minkowski space with radii $r_-< r_+$.  The imaginary part of this unstable mode is plotted in figure~\ref{fig:TwoCavities}.  We see again that each mode is periodically unstable, but no unstable region remains when we consider all modes.

\subsection{dS}
\label{sec:dsSphere}

Our final background is de Sitter ($K=1$, $M=0$, $\lambda > 0$ in \eqref{Schw background}). The dS radius $L_{dS}$ is given by $L_{dS} = \lambda^{-1/2}$ and we restrict ourselves to cases in which the Dirichlet wall fits inside the cosmological horizon,
$r_D < L_{dS}$.  This makes the tensor and vector potentials positive definite.

\subsubsection{Inside}
We begin with the interior region $0 \le r < r_D$. Solutions that are regular at the origin can be written
\begin{align}
	\phi^{(T)} &= \rho^{\ell_T + n/2} (1 - \rho^2)^{i \tilde \omega/2} \hs _2 F_1 \left[ \frac{1}{2}(\ell_T + i \tilde \omega),
	\frac{1}{2} ( \ell_T + i \tilde \omega + n + 1 ); \frac{1}{2} (2 \ell_T + n+1); \rho^2  \right], \\
\label{phi V dS inside}
	\phi^{(V)} &= \rho^{\ell_V + n/2} (1 - \rho^2)^{i \tilde \omega/2} \hs _2 F_1 \left[ \frac{1}{2} (\ell_V + i \tilde \omega + 1),
	\frac{1}{2}(\ell_T + i \tilde \omega + n ); \frac{1}{2} ( 2 \ell_V + n+1 ); \rho^2  \right], \\
	\phi^{(S)} &= \rho^{\ell_S + n/2} (1 - \rho^2)^{i \tilde \omega/2} \hs _2 F_1 \left[ \frac{1}{2} (\ell_S + i \tilde \omega + 2),
	\frac{1}{2}(\ell_S + i \tilde \omega + n -1); \frac{1}{2} ( 2 \ell_S + n+1 ); \rho^2  \right],
\end{align}
\noindent where $\rho = r/L_{dS}$ and $\tilde \omega = \omega L_{dS}$.
The spectra are again given by imposing the boundary conditions \eqref{bc MF tensor}, \eqref{bc MF vector}, \eqref{bc MF scalar}
at $\rho = \rho_D = r_D/L_{dS}$. As usual,  we
can readily adapt the proof of \cite{Ishibashi:2003ap} to forbid both tensor and vector instabilities.
On the other hand, we do find instabilities in the scalar sector when the wall is close enough to the cosmological horizon.
These instabilities can be understood by looking at \eqref{LSP vector scalar} more closely.
First, note that the coefficients characterizing the boundary condition \eqref{gen bc at wall} can be obtained from
\eqref{aS AdS}, \eqref{bS AdS}, \eqref{cS AdS} by the analytic continuation $L_{dS} = i L_{AdS}$. By doing so, we see that
$b_S$ is no longer positive definite, so our usual stability proof fails. It is natural to expect that the critical value $r_{crit}$
for which $b_S(r_{crit}) = 0$ to be an upper bound such that there are no instabilities for $r_D < r_{crit}$.
This turns out to be the case. For $n = 2$, $r_{crit}$ is independent of $\ell_S$, with $r_{crit}/L_{AdS} = 1/\sqrt{2}$. For higher values of $n$,
the value of $r_{crit}$ depends on $\ell_S$ in a somewhat complicated way, but at large
$\ell_S$ we have
\begin{equation}
	r_{crit}/L_{AdS} = \sqrt{\frac{n-1}{n} } + O(\ell_S^{-2}).
\end{equation}

\begin{figure}[h]
\center
\subfigure[][]{
\includegraphics[width=0.4\linewidth]{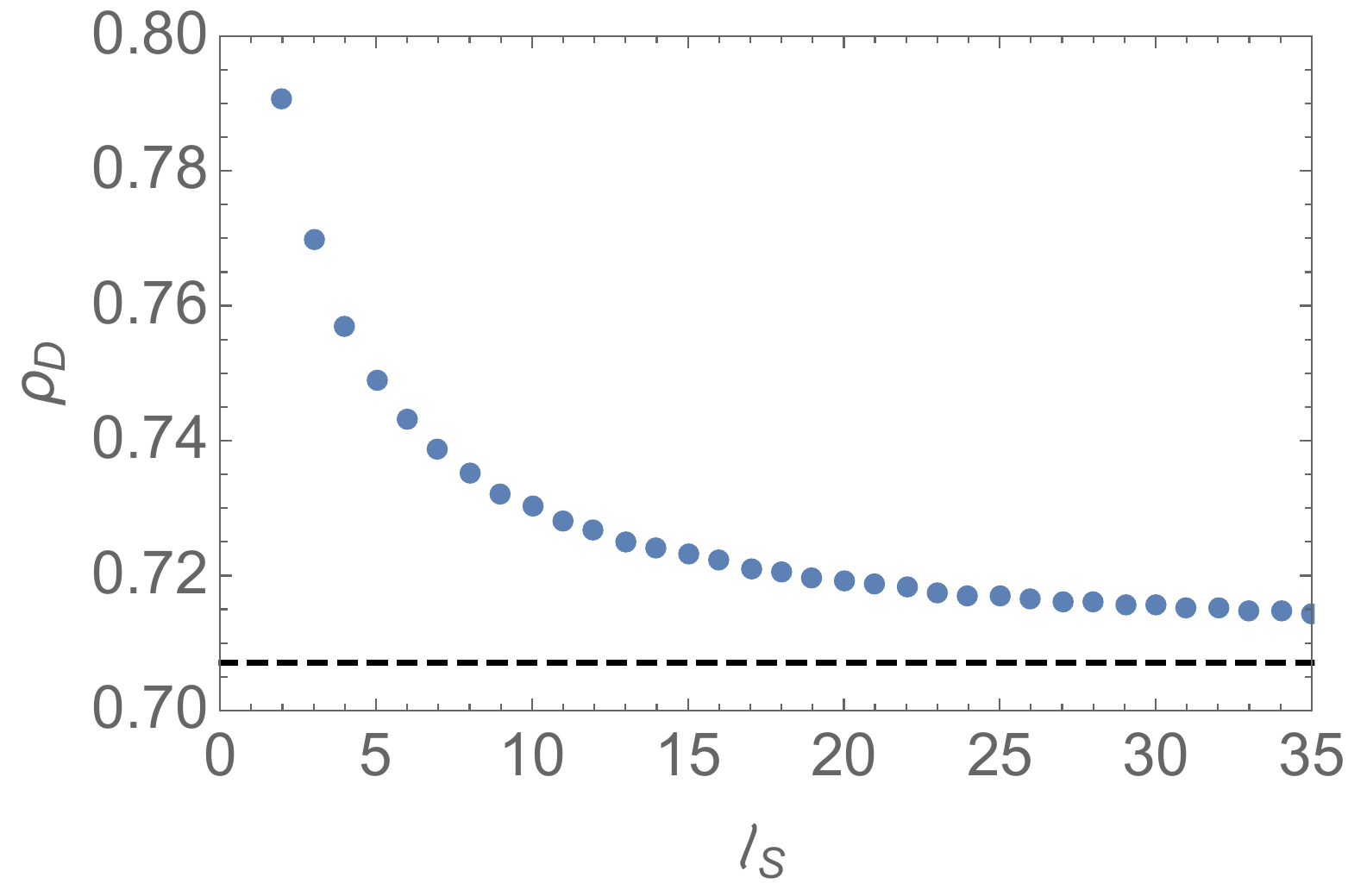}
}\qquad\qquad
\subfigure[][]{
\includegraphics[width=0.40\linewidth]{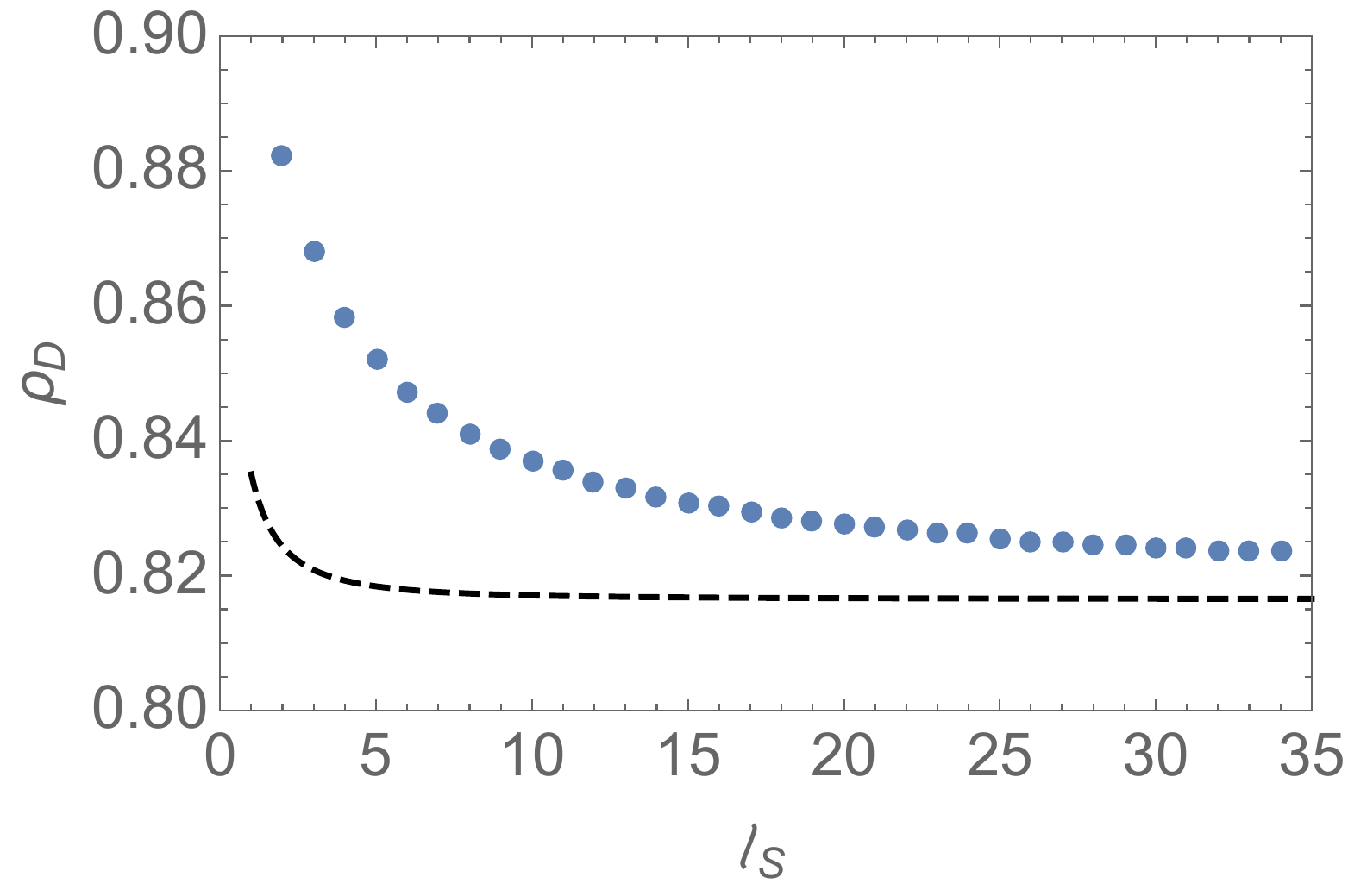}
}
\caption{{\bf Unstable internal dS scalar modes:} The data points are the value of the location of the wall $\rho_D$
at which the first unstable modes appear as a function of angular momentum $\ell_S$, while the dashed line is $r_{crit}/L_{AdS}$. Results are shown for $n=2$ (left) and $n=3$ (right).}
\label{inst dS in}
\end{figure}

Figure \ref{inst dS in} plots the value of $\rho_D$ at which the first unstable mode appears as a function of $\ell_S$.

This instability is particularly interesting both because it is the only instability we find in the interior of a cavity and because it is connected to the fluid/gravity analysis of~\cite{Bredberg:2010ky,Bredberg:2011jq}.  As shown in~\cite{Bredberg:2010ky}, the limit $r_D\to L_\text{dS}$ reduces the dynamics of the gravitational field to that of an incompressible fluid on the Dirichlet surface.  The stress tensor of this fluid is given by the Brown--York expression \cite{Brown:1992br} evaluated on the wall; i.e., by
\begin{align}
\label{tBY}
t_{AB} = -\frac{1}{8\pi G} \left( K_{AB} - K \gamma_{AB} - C \gamma_{AB}\right) \, ,
\end{align}
where $\gamma_{AB}$ is the induced metric on the Dirichlet wall, $K_{AB}$ is the extrinsic curvature, and $C$ is an arbitrary constant that fixes the zero point of the energy.  According to this correspondence the energy density of the fluid is given by
\begin{align} \label{eq:BY00}
-t_0{}^0 = \frac{1}{8\pi G}  (K_0{}^0 - K \gamma_0{}^0 - C \gamma_0{}^0) = \frac{1}{8\pi G} \sqrt{\gamma}  (K_I{}^I -C) \, ,
\end{align}
where $I$ runs over all of the boundary indices except $t$.  Unlike the explicit examples considered in~\cite{Bredberg:2010ky}, in the present case the size of the spheres increase as we move from the cutoff surface towards the cosmological horizon, which has the effect of changing the sign of the $K_I{}^I$ term in~\eqref{eq:BY00}.  This means that the fluid has a negative energy density or, if $C$ is chosen to be sufficiently negative, a term with negative pressure.  Given this unconventional sign it is plausible that the fluid is unstable, though it would be interesting to develop a more complete understanding of the fluid analog of this gravitational instability.

\subsubsection{Outside}

\begin{figure}[h]
\center
\subfigure[][]{
\includegraphics[width=0.4\linewidth]{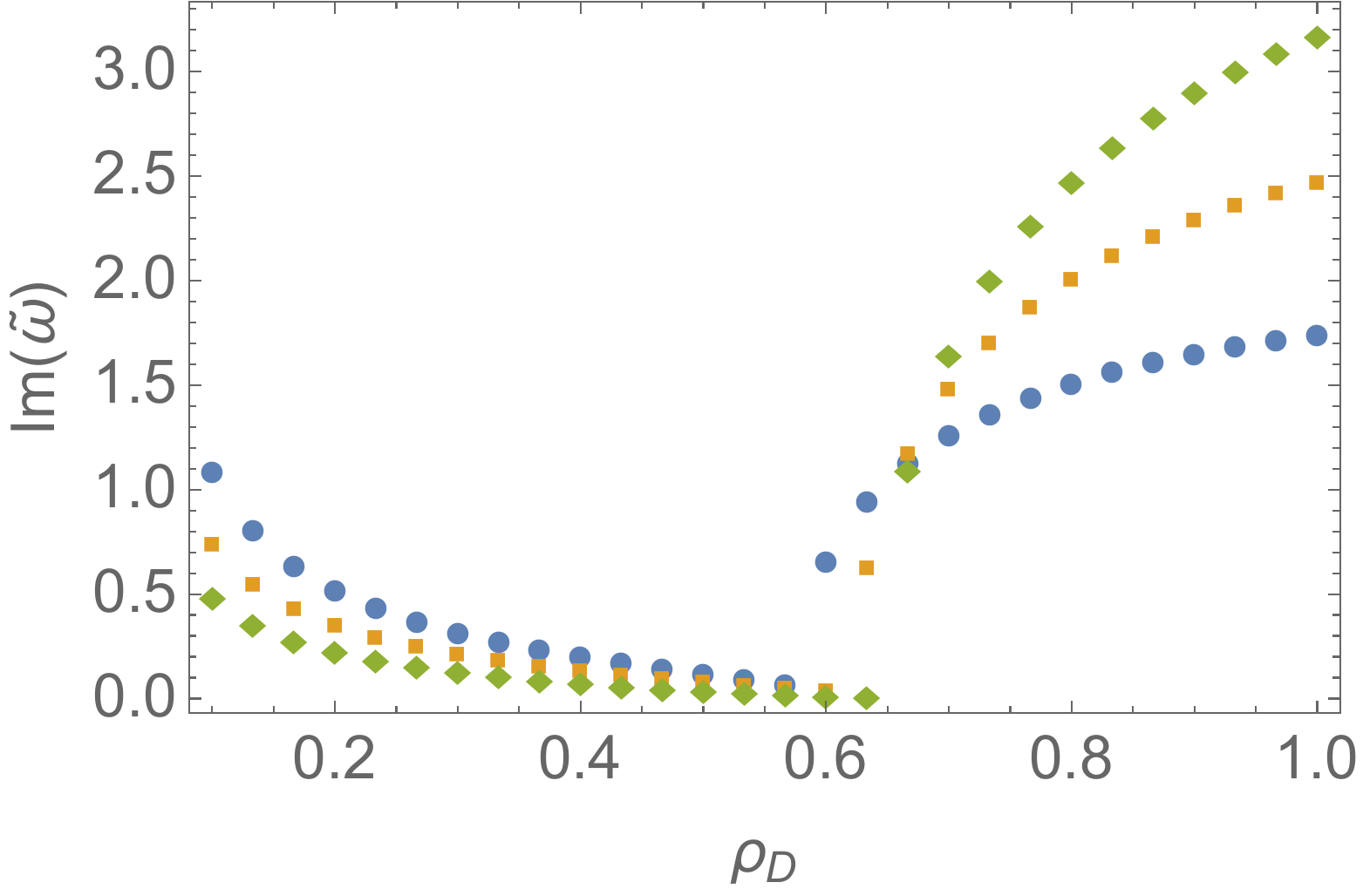}
}\qquad\qquad
\subfigure[][]{
\includegraphics[width=0.40\linewidth]{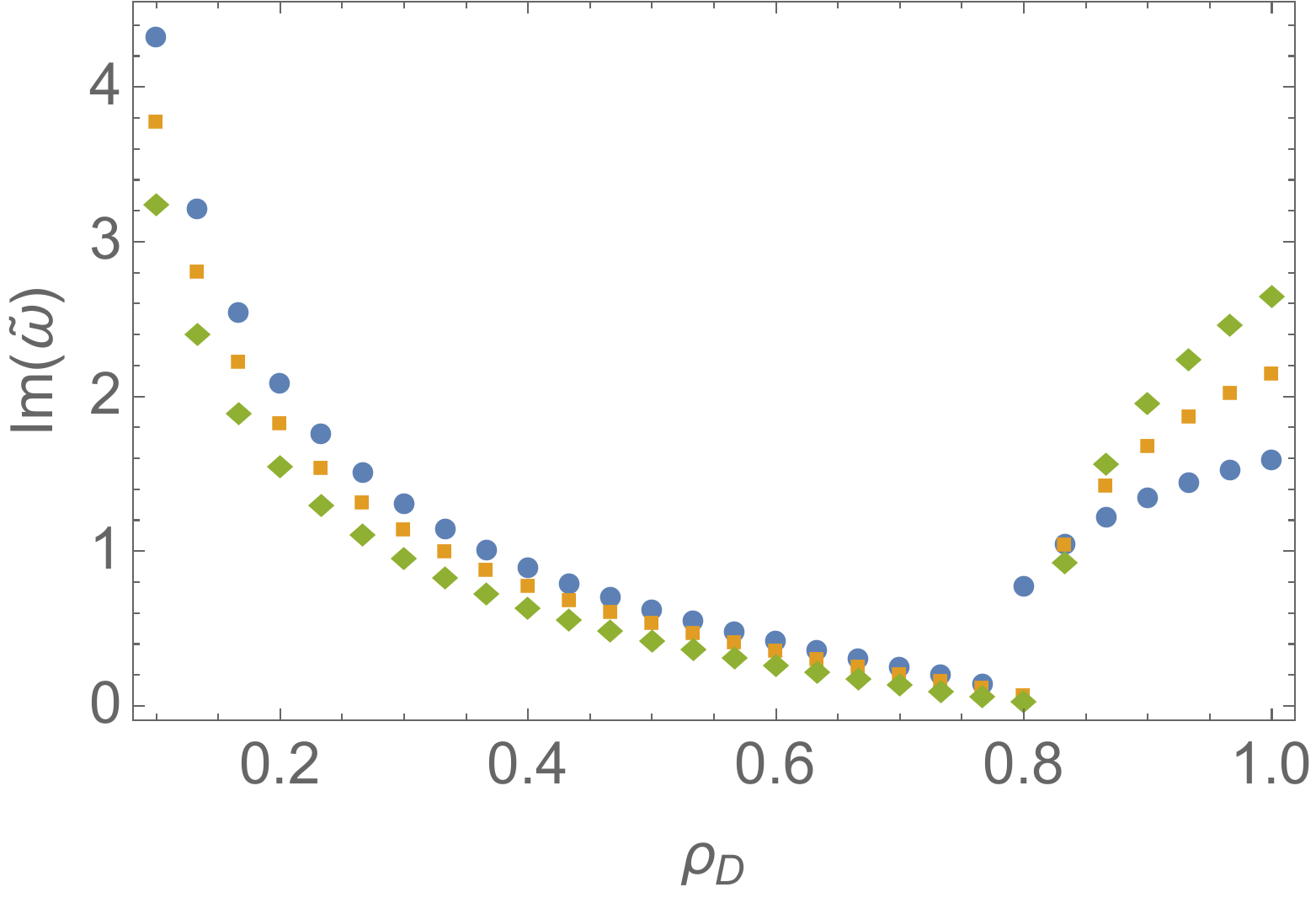}
}
\caption{{\bf Unstable external dS scalar modes:} Value of the imaginary part of the unstable modes as a function of $\rho_D$,
for $\ell_S =2,3,4$ (blue circles, yellow squares, green diamonds), with $n=2$ (left) and $n=4$ (right). For small $\rho_D$,
the instabilities for each $\ell_S$ are a pair of complex modes in the upper half plane. As we increase $\rho_D$, ${\rm Im} \, \tilde \omega $
and ${\rm Re} \, \tilde \omega$ decrease until these modes merge on the positive imaginary axis, after which
${\rm Im} \, \tilde \omega$ increases approaching a finite value as $\rho_D \to 1$.}
\label{inst dS out}
\end{figure}

Turning to the exterior modes, we have $r \in (r_D, L_{dS})$, or equivalently, $\rho \in (\rho_D, 1)$. We impose outgoing boundary
conditions at the cosmological horizon, so that the profiles behave as $\phi^{(I)} \sim (1-\rho)^{- i \tilde \omega/2}$ near $\rho=1$.
The corresponding solutions are
\begin{align}
\label{dS out profile T}
	\phi^{(T)} &= \rho^{\ell_T + n/2} (1 - \rho^2)^{- i \tilde \omega/2} \hs
	_2 F_1 \left[ \frac{1}{2}(\ell_T - i \tilde \omega), \frac{1}{2}(n+1 + \ell_T - i \tilde \omega); 1 - i \tilde \omega;
	1 - \rho^2 \right] \\
\label{dS out profile V}
	\phi^{(V)} &= \rho^{\ell_V + n/2} (1 - \rho^2)^{- i \tilde \omega/2} \hs
	_2 F_1 \left[ \frac{1}{2}(1+ \ell_V - i \tilde \omega), \frac{1}{2}(n + \ell_V - i \tilde \omega); 1 - i \tilde \omega;
	1 - \rho^2 \right] \\
\label{dS out profile S}
 	\phi^{(S)} &= \rho^{\ell_S + n/2} (1 - \rho^2)^{- i \tilde \omega/2} \hs
	_2 F_1 \left[ \frac{1}{2}(2 + \ell_S - i \tilde \omega), \frac{1}{2}(n-1 + \ell_S - i \tilde \omega); 1 - i \tilde \omega;
	1 - \rho^2 \right]
\end{align}
Imposing boundary conditions at the wall, we find no unstable modes in the tensor and vector sectors (as usual, the tensor can be proven to be stable as in section \ref{MinkowskiOutside}). For the
scalars, however, we find instabilities for all $\rho_D < 1$, see figure \ref{inst dS out}.

\section{Results for flat walls}
\label{sec: flat wall results}

We now consider static spacetimes with planar Dirichlet walls.  Thus we set $K=0$  in \eqref{Schw background} (with $M=0$ as usual). The restriction to static such spacetimes limits our backgrounds to empty Minkowski space and Poincar\'e AdS.
Recall that for $K=0$ the spectrum of transverse momenta $k_I^2$ is non-negative and continuous.

In all cases we will find the system to be linearly stable.  With the given planar symmetry, we find it convenient to simply calculate QNM frequencies analytically rather than give an abstract argument for stability.

\subsection{Minkowski space}

We first study fluctuations in half of Minkowski space in the presence of a flat Dirichlet wall. Due to the symmetries
of the problem, it is more convenient to use a Cartesian coordinate system rather than the one in \eqref{Schw background}, so our
discussion in this section largely departs from the Kodama-Ishibashi formalism.
We denote one of the spatial coordinates by $z$ and place the Dirichlet wall at $z = 0$.
To make this choice explicit we write the line element as
\begin{equation}
	ds^2 = \eta_{A B} dy^A dy^B + dz^2,
\end{equation}
\noindent where $y^A$ includes the time coordinate $t$ and the remaining $n$ spatial coordinates. We consider a perturbation
$h_{\mu \nu}$, upon which we demand $h_{A B}|_{z=0} = 0$.  It is convenient to impose the following gauge condition
\begin{align} \label{eq:MinkGauge}
\partial^\mu  \bar h_{\mu\nu}   = 0
\end{align}
where
\begin{align}
\bar h_{\mu\nu} := h_{\mu\nu} - \frac{1}{2} h \, \eta_{\mu\nu} \ .
\end{align}
To demonstrate that this gauge choice is consistent with our boundary condition it is sufficient to note that an arbitrary perturbation can be put into the gauge~\eqref{eq:MinkGauge} by a linearized diffeomorphism generated by the vector $\xi$ defined as the solution to the initial boundary value problem\footnote{A solution can always be constructed by appropriately extending the problem to an entire Cauchy surface of Minkowski space, choosing parity odd initial data, time evolving, and then only keeping the solution in the region $z\ge 0$.}
\begin{align}
\partial^2  \xi_\nu = -\partial^\mu  \bar  h_{\mu\nu}  \qquad \left.  \xi^\mu \right|_{z=0} = 0 \ .
\end{align}
In this gauge the field equation takes the from $\partial^2 \bar h_{\mu\nu} = 0$.  Using translations invariance of the background spacetime in the $y^A$ directions we can make the Fourier decomposition
\begin{align} \label{eq:hFourier}
\bar h_{\mu\nu} = \int dk \, \bar h_{\mu\nu}^{(k)}(z)  e^{-i (\omega t+ k_I y^I)}    \ ,
\end{align}
where $I$ runs over all of the $y^A$ except $t$.

We will now prove stability by showing that $\omega $ is real.  Inserting~\eqref{eq:hFourier} into the field equation $\partial^2 \bar h_{\mu\nu} = 0$ gives
\begin{align}
(\partial_z^2 +\omega ^2 -  k_I k^I ) \bar h_{\mu\nu}^{(k)}(z) = 0 \ ,
\end{align}
which has the solution
\begin{align}
\bar h_{\mu\nu}^{(k)}(z) = s_{\mu\nu}^{(k)} \sin\left(k_z z\right) + c_{\mu\nu}^{(k)} \cos\left(k_z z\right) \qquad k_z := \sqrt{\omega^2 - k_I k^I} \, ,
\end{align}
where $s_{\mu\nu}^{(k)},c_{\mu\nu}^{(k)}$ are constants and our boundary condition requires that
\begin{align}
c_{AB}^{(k)} - \frac{1}{2} c^{(k)} \eta_{AB}=0 \, .
\end{align}
We wish to construct asymptotically flat solutions out of a superposition of these modes.  In order to to do so we must require that the modes be plane wave normalizable.  This implies that the $k_I$ and $k_z $ are real.  Therefore
\begin{align}
\omega^2 \ge k_I k^I \ge 0 \ .
\end{align}
Furthermore if $\omega= 0$ then the components of $h_{\mu\nu}$ are constants and only the $h_{z\mu}$ are non-vanishing.   This solution is equivalent to a trivial rescaling of the coordinates.  Therefore Minkowski space with a Dirichlet wall is linearly stable to gravitational fluctuations.

\subsection{AdS}

We now consider the Poincare patch of AdS ($K=0$, $M=0$ and $\lambda < 0$ in
\eqref{Schw background}). For convenience we introduce the dimensionless radial coordinate $\rho = r /L_{AdS}$.
We mention that scalar fields in the Poincare patch of AdS with various geometrical cutoffs and boundary conditions were previously studied in \cite{Andrade:2011aa}.

Any mode in Poincar\'e AdS can be classified by its ``boundary mass'' defined through
$m^2_{bndy, I} = \omega^2 - k_I^2$. We refer to the modes with  $m^2_{bndy, I} > 0$ as time-like, those for which
$m^2_{bndy, I} = 0$ as null and those for which $m^2_{bndy, I} <0$ as space-like or tachyonic. In this last case $\omega^2$
is not positive definite, so space-like modes imply instabilities. In order to establish stability, we only need to rule out these tachyonic
modes and we will thus concentrate on these excitations.
Unlike Minkowski space where both sides of a flat wall are identical, flat walls in Poincar\'e AdS admit a natural notion of inside the wall vs. outside.

\subsubsection{Inside}

We begin with the region $0 < \rho < \rho_D$ inside the wall.  As mentioned above, it suffices to restrict our attention
to the tachyonic modes $m^2_{bndy, I} <0$. These modes decay exponentially near the Poincar\'e horizon, so the reasoning
leading to \eqref{LSP vector scalar} applies. The boundary conditions at the wall for vectors and tensors are the same as those
in Minkowski space, which implies $\omega^2 > 0$, in contradiction with our original assumptions of $m^2_{bndy, I} <0$. Therefore,
we conclude that there are no instabilities in the tensor and vector sectors.

For the scalars, the potential reduces to
\begin{equation}
	V_S = L_{AdS}^{-2} \left( \frac{k_S^2}{r^2} + \frac{(n-2)(n-4)}{4} \right)
\end{equation}
\noindent which is positive definite for $n\neq 3$. The boundary condition takes the form \eqref{gen bc at wall} with
\begin{equation}
a_S = (n-1) r^3, \qquad b_S =  \frac{n}{2} (n-2)(n-1) r^2 +  k_S^2 L_{AdS}^2 , \qquad c_S = L^4_{AdS}
\end{equation}
These are all positive, so that using \eqref{LSP vector scalar} we arrive at $\omega \geq 0$ which once again rules out the
space-like modes.
Since the potential is not positive for $n=3$, we need to resort to a more detailed analysis in this case.
To simplify the notation, we introduce $\kappa^2_S = - m^2_{bndy, S}$ and take ${\rm Re} \, \kappa_S > 0$.
The tachyonic profile for the scalar modes is given by
\begin{equation}
\label{phi S PAdS}
	\phi^{(S)} = \rho^{-1/2} K_{(n-3)/2}(\kappa_S/\rho).
\end{equation}
\noindent where $K_\nu(x)$ is a Bessel function of the second kind.  Replacing this into the boundary conditions
at the wall \eqref{bc MF scalar} we find
\begin{equation}
  		K_{(n+1)/2}(x_S) = 0
\end{equation}
\noindent where $x_S = \kappa_S/\rho_D$. It is well-known (see e.g. \cite{abramowitz+stegun}) that this equation has no solutions
in the complex plane with ${\rm Re} \, x_S >0$. It follows that there are no tachyons in the spectra, and that the interior region of Poincare AdS remains linearly stable against gravitational perturbations in the presence of a Dirichlet wall.

\subsubsection{Outside}

Consider now $\rho \in (\rho_D, \infty)$ and introduce the radial variable $u = 1/\rho$.
As for $K=1$ case, the boundary condition at $u=0$ fixes the conformal metric, which amounts to
imposing \eqref{bc MF tensor}, \eqref{bc MF vector}, \eqref{bc MF scalar} in the limit $u_D \rightarrow 0$. The solutions that satisfy this condition are
\begin{align}
\label{phi T PAdS out}
	\phi^{(T)} &= u^{1/2} J_{(n+1)/2}(m_{bndy, T} u), \\
\label{phi V PAdS out}
	\phi^{(V)} &= u^{1/2} J_{(n-1)/2}(m_{bndy, V} u), \\
\label{phi S PAdS out}
	\phi^{(S)} &= u^{1/2} J_{(n-3)/2}(m_{bndy, S} u).
\end{align}
Once again, the boundary conditions \eqref{bc MF tensor}, \eqref{bc MF vector}, \eqref{bc MF scalar} at the wall lead to identical results in all three sectors, namely
\begin{equation}\label{J zeroes}
 		J_{(n+1)/2} (x_I) = 0,
\end{equation}
\noindent where $x_I = m_{bndy I} u_D$ with $u_D = 1/\rho_D$. The zeroes of this equation are real \cite{abramowitz+stegun}
so all solutions have $m_{bndy I}^2 > 0$; there are no instabilities.

\section{Discussion}
\label{sec:discussion}

Our work above studied the stability of vacuum gravity solutions under linear perturbations
in the presence of a Dirichlet wall. We considered flat or spherical walls in Minkowski and Anti-de Sitter space, and spherical walls in de Sitter.  Stability was analyzed in the regions both inside and outside such walls.
We also studied the addition of a black hole to the region inside the wall.  For generic modes the dissipative horizon only increases stability (see Appendix~\ref{app:BlackHoles}), but as discussed in section \ref{sec:l=1 modes}, adding the black hole also leads to new ``non-generic'' modes associated with moving the black hole center of mass relative to the wall.    The stability of these modes (for $\lambda \le 0$, and for $\lambda >0$ when the black hole horizon is sufficiently close to the wall) provide evidence that neutral black holes are repelled from Dirichlet walls as predicted by \cite{Andrade2015moduli}. \\

Our investigation of generic modes relied heavily on the Kodama-Ishibashi formalism, which reduced the problem to decoupled
ODEs obeyed by three master fields (tensor, vector,
and scalar).
For tensor and vector modes the master field equations define standard eigenvalue problems for the (quasi-)normal modes frequencies.
In particular, the Dirichlet boundary condition on the metric forces the tensor master field to vanish at the wall, while it imposes a Robin (or mixed type) condition on the vector master field.  As one would expect from analogy with spin-0 and spin-1 fields, we found no instability in either of these sectors for any of the cases studied.  However, the scalar sector satisfied a more general boundary condition with explicit dependence on the frequency. This feature gave rise to instabilites that would not occur for fields of spins zero or one.
\\

Our main result was to identify when instabilities actually occur.  The examples with flat Dirichlet walls were always stable.  And inside spherical Dirichlet walls the scalar sector was unstable only with a positive cosmological constant.  This instability can apparently be understood by noting that the coefficient $b_S$ in \eqref{gen bc at wall} becomes negative.  Indeed, as $\ell_S \rightarrow \infty$ our plots (figure \ref{inst dS in}) of the onset of instability for each $\ell_S$ appear to approach the locus where $b_S$ changes sign. On the other hand, the field outside a spherical cavity was unstable in all cases considered.  In the AdS case, the stability of individual modes displayed an interesting pattern that also appears when one studies the cavity between two concentric spherical walls in Minkowski space; see figures \ref{fig:AdS islands} and \ref{fig:TwoCavities}.

It is tempting to associate the unstable modes outside spherical cavities with the fact that the Dirichlet wall allows the ADM energy boundary term at infinity to be arbitrarily negative.  In particular, the boundary conditions defined by any spherical Dirichlet wall admit negative-mass (perhaps AdS or dS) Schwarzschild solutions with arbitrary $M <0$.  The wall hides the naked singularity and the solution is smooth in the physical region outside the wall.   However, the same negative ADM energy would arise in planar AdS where the system is stable.  Furthermore, we note that the system outside the wall has {\it two}  boundaries -- the asymptotic one at infinity and the boundary at the wall itself.  As a result, time-translations along the wall are not gauge transformations and instead define an additional class of global symmetries\footnote{Standard terminology (see e.g. \cite{Fischetti:2012rd}) would refer to them as asymptotic symmetries, but this choice of words could cause confusion for a diffeomorphism supported near a Dirichlet wall at finite distance.} associated with a separately-conserved energy $H_{wall}$ given by an appropriate integral of the Brown-York stress tensor \eqref{tBY}.  The structure is similar to that of a two-sided wormhole where the ADM Energy is separately conserved in each asymptotic region. On (perhaps AdS or dS) Schwarschild, extending the computations of \cite{Brown:1992br} to $\lambda \neq 0$ and using the sign appropriate to being outside the wall gives
\begin{equation}
H_{wall} = \frac{n}{2} r_D \sqrt{f(r_D)}= \frac{n}{2} r_D \sqrt{K -\frac{2M}{r_D^{n-1}} - \lambda r^2_D}.
\end{equation}
This wall-energy is minimized by taking $M$ sufficiently positive so that $r_H = r_D$, while $H_{wall}$ becomes large and positive as $M \rightarrow -\infty$.  In particular, conservation of $H_{wall}$ seems to exclude what would otherwise seem to be a natural scenario in which initially-flat Minkowski space outside the wall decays to $M < 0$ Schwarzschild with positive-energy radiation far from the wall.   The ultimate fate of our instability thus remains an interesting open question.

In a recent article \cite{Andrade2015moduli}, we used the small velocity approximation to study the motion of a spherical extreme electrically-charged black hole in an asymptotically flat spacetime that contains a flat Dirichlet wall.  Our results suggested that system to be
unstable due to the appearance of negative kinetic energies for black holes near the wall.  But at least at the linear stability level, the present investigation of vacuum solutions found no sign of linear instabilities for
flat Dirichlet walls.  Indeed, as noted above, neutral black holes seem to be repelled by Dirichlet walls.  It would thus be interesting to use the master field formalism of  \cite{Kodama:2003kk} to study the linear stability of
charged solutions in the presence of Dirichlet walls.  It would also be interesting to study these problems at the truly non-linear level.  We leave such questions for future work.

The nonlinear stability of Dirichlet-wall backgrounds with respect to spherically symmetric spin-0 perturbations has received much attention over the past few years \cite{Maliborski:2012gx,Okawa:2014nea}. In \cite{Maliborski:2012gx,Okawa:2014nea}, a nonlinear instability was found for arbitrarily small spherically symmetric spin-0 perturbations. The origin of this instability has been attributed to the fact that spherically symmetric spin-0 perturbations have a fully resonant spectrum at the linear level\footnote{This means that there exists an infinite number of tetrads of normal mode frequencies satisfying $\pm \omega_1\pm\omega_2\pm\omega_3=\pm\omega_4$.}~\cite{Bizon:2011gg,Dias:2011ss,Dias:2012tq,Maliborski:2014rma}. But when these systems are only asymptotically resonant at large $\omega$, they can be nonlinearly stable \cite{Dias:2012tq}. One may check that with zero or negative cosmological constant our spin-2 perturbations inside spherical Dirichlet-walls approach resonance even more slowly as $\omega \rightarrow \infty$. We therefore expect these Dirichlet-wall systems to be nonlinearly stable as well.

\acknowledgments

It is a pleasure to thank Gary Horowitz and Helvi Witek for helpful discussions and feedback.
T.A. was supported by the European Research Council under the European Union's Seventh Framework Programme
(ERC Grant agreement 307955).
W.K. and D.M. were supported by the National Science Foundation under grant number PHY12-05500 and by funds
from the University of California.
T.A. thanks the Galileo Galilei Institute for Theoretical Physics for their hospitality during the final stages of this work.
D. M. also thanks the KITP for their hospitality during the final stages of the project, where his work was also supported in part by National Science foundation grant number PHY11-25915.

\appendix

\section{Symplectic structure for Dirichlet boundary conditions}
\label{app:symplectic}

This appendix studies the symplectic structure for the theory
defined with the Dirichlet wall in order to determine which diffeomorphism-modes ($\ell_S =1$) are non-trivial and which are pure gauge.
We note that the symplectic current receives
a contribution from the Gibbons-Hawking term which plays a crucial role in its conservation -- unless one works in radial gauge where this contribution vanishes.
After constructing an appropriate inner product, we compute the norms of the diffeomorphism-modes described in section
\ref{sec:l=1 modes}.

We begin by recalling the algorithm of \cite{Compere:2008us} for constructing a conserved symplectic structure from a well-defined variational principle for a field theory in the presence of a boundary.  See also \cite{Lee:1990nz,Iyer:1994ys,Wald:1999wa} for related treatments of covariant phase spaces which do not study such boundaries in detail.

Denoting the (not necessariy scalar) fields by $\phi$, we assume that the action
\begin{equation}
	S[\phi] = \int_M L_0 + \int_{\partial M} L_{\partial}
\end{equation}
\noindent has an extremum when some boundary condition $b(\phi) =0$ is satisfied.  This $b$ can be any local functional of the fields $\phi$.  Thus
\begin{equation}\label{Dbc condition app}
	\delta S = \int_{\partial M} \pi_b \delta b
\end{equation}
\noindent when the bulk equations of motion are satisfied, where $\pi_b$ may be called the momentum conjugate to $b$. As usual, we take $\partial M$ to be the part of the boundary where boundary conditions need to be imposed in order to define a phase space. In particular, we neglect any terms lying at past or future boundaries of the system.

Writing the variation of the bulk term as
\begin{equation}
	\delta L_0 = ({\rm eoms}) \delta \phi + d \theta_0,
\end{equation}
\noindent the condition \eqref{Dbc condition app} requires the pull-back of $\theta_0$ to $\partial M$ to satisfy
\begin{equation}
\label{eq:sympkey}
	\theta_0|_{\partial M} = \pi_b \delta b - \delta L_\partial + d \theta_\partial
\end{equation}
for some $\theta_\partial$.  The total derivative $d \theta_\partial$ does not contribute to \eqref{Dbc condition app} since we again neglect terms lying at any past or future boundaries.
As argued in \cite{Compere:2008us}, we obtain a conserved symplectic structure by taking the symplectic current to be
\begin{equation}
 	j = j_0 - d j_\partial,
 \end{equation}
\noindent where $j_0$ and $j_\partial$ are the symplectic currents associated to the potentials
$\theta_0$ and $\theta_\partial$, i.e.
\begin{equation}
	j_0 = \delta_2 \theta_0[\delta_1 \phi] - \delta_1 \theta_0[\delta_2 \phi], \qquad
	j_\partial = \delta_2 \theta_\partial[\delta_1 \phi] - \delta_1 \theta_\partial[\delta_2 \phi].
\end{equation}
The key point is that, since the anti-symmetric second variation of $L_\partial$ vanishes identically, taking the anti-symmeric variation of \eqref{eq:sympkey} implies that $j$ vanishes when pulled back to $\partial M$ ($j|_{\partial M} =0$) and evaluated on variations satisfying the desired boundary condition (so that $\delta b=0$).  Since no symplectic current flows though the boundary, conservation of the symplectic structure $\int_{\Sigma} j$ for hypersurfaces $\Sigma$ having boundaries only on $\partial M$ follows immediately from the fact that the bulk contribution to this current is closed \cite{Lee:1990nz,Iyer:1994ys,Wald:1999wa}, $d j_0 = 0$.

We now apply this construction to gravity with Dirichlet boundary conditions. We are interested in
space-times for which $\partial M$ is a time-like surface of constant $r$,
with unit normal $n_\mu$.
We assume that they can be globally foliated by constant $r$ surfaces, on which we introduce coordinates $y^i$,
so that the metric can be written in the form
\begin{equation}
	ds^2 = N^2 dr^2 + \gamma_{ij} (dy^i + N^i dr) (dy^j + N^j dr),
\end{equation}
\noindent where the induced metric on surfaces of constant $r$ is $\gamma_{ij}$ and $N$, $N^i$ are the radial lapse and shift
functions, respectively.
Note that the normal satisfies $n_\mu dx^\mu = N dr$.

For Dirichlet boundary conditions on $\partial M$, a suitable action is given by adding the Gibbons-Hawking term to the bulk
Einstein-Hilbert term
\begin{equation}\label{total action app}
 	S = \int_M \sqrt{g} (R - 2 \Lambda) + 2 \int_{\partial M} \sqrt{\gamma} K,
\end{equation}
\noindent where $\gamma_{\mu \nu} = g_{\mu \nu} - n_\mu n_\nu$ is the induced metric at the boundary
and $K$ is the trace of the extrinsic curvature $K_{\mu \nu} = \gamma_\mu \hs^\sigma \nabla_\sigma n_\nu$.
In this covariant notation, tensors on $\partial M$ are degenerate space-time tensors which vanish when contracted with $n_\mu$.
In particular, note that $\gamma = {\rm det} \, \gamma_{ij}$ and $\gamma \neq {\rm det} \, \gamma_{\mu \nu} = 0$. \\

The bulk contribution to the symplectic current is the standard one for Einstein-Hilbert gravity, which we take to be given by
\begin{equation}\label{j EH}
	j^\nu_{EH} = \delta_2 ( \sqrt{g} g^{\alpha \beta} ) \delta_1 \Gamma^\nu _{\alpha \beta} - \delta_2 ( \sqrt{g} g^{\alpha \nu} ) \delta_1 \Gamma^\beta_{\alpha \beta}
	- (1 \leftrightarrow 2).
\end{equation}
See \cite{Barnich:2001jy, Barnich:2007bf} for other choices of symplectic currents that differ from \eqref{j EH} by total derivatives.
A general on-shell variation of the action \eqref{total action app} is of the form
\begin{equation}
	\delta S = \int_{\partial M} \sqrt{\gamma} ( \pi^{\mu \nu} \delta \gamma_{\mu \nu} + {\cal D}_\mu c^\mu),
\end{equation}
\noindent where ${\cal D}_\mu$ is the covariant derivative compatible with $\gamma_{\mu \nu}$, the conjugate momentum is given by
\begin{equation}
	\pi^{\mu \nu} = - ( K^{\mu \nu} - K \gamma^{\mu \nu} ),
\end{equation}
\noindent and
\begin{equation}
 	c^\mu = - \gamma^{\mu \rho} \delta g_{\rho \sigma} n^\sigma
 \end{equation}
is tangent to $\partial M$ so that ${\cal D}_\mu c^\mu$ is well-defined.

Since our boundary conditions are $\delta \gamma_{\mu \nu} |_{\partial M} = 0$, the boundary contribution to the symplectic potential
is
\begin{equation}
	\theta_\partial^i = - \sqrt{\gamma} g^{i \lambda}  \delta g_{\lambda \sigma} n^\sigma,
\end{equation}
The boundary contribution to the symplectic current is obtained by taking the antisymmetrized variation
\begin{equation}
	j_\partial^i = - \delta_2 (\sqrt{\gamma} g^{i \lambda} n^\sigma) \delta_1 g_{\lambda \sigma} + (1 \leftrightarrow 2),
\end{equation}
and the total symplectic structure is
\begin{equation}\label{Omega full}
	\Omega = \int_\Sigma j_{EH} - \int_{\partial \Sigma} j_\partial .
\end{equation}
As noted above, general arguments imply that it is conserved.  We have also checked this result by direct computation.

The combination
\begin{equation}
	(\delta_1 g, \delta_2 g) = - i \Omega(\delta_1 g, \delta_2 g^*)
\end{equation}
\noindent defines the desired inner product, where $^*$ denotes complex conjugation. We now explicitly evaluate this inner product for the diffeomorphism modes discussed in section \ref{sec:l=1 modes}.  For simplicity we consider only the case of four spacetime dimensions ($n=2$), though other dimensions should behave similarly.

It is well known that the bulk contribution to the symplectic structure can be expressed as a total derivative
for all diffeomorphims, allowing us to reduce the radial integral to boundary terms.
We focus on the contribution from the Dirichlet wall at $r_D$, since we can always choose our diffeomorphism to vanish in the neighborhood of any other boundary.
Due to the orthogonality of the spherical harmonics, our diffeomorphism-modes can have non-trivial inner products only with other $\ell_S = 1$ modes; i.e., only with other pure diffeomorphisms.

Recall from section \ref{sec:l=1 modes} that the frequencies are either purely real or purely imaginary.
For the real case, which occurs when $f'(r_D) >0$, we find
\begin{equation}\label{ip l=1}
	(\delta_1 g, \delta_2 g) = \frac{6 M \omega_{\ell_S = 1}}{\pi f(r_D)}   L_1(r_D) L_2^*(r_D),
\end{equation}
where we remind the reader that $L$ is the free function controlling the radial dependence of the diffeomorphism.
The result is non-vanishing for $L(r_D) \neq 0$, so such excitations are physical as expected.
Conversely, modes with $L(r_D) = 0$ have vanishing inner product with all modes and therefore are pure gauge.

The imaginary frequencies are similar\footnote{A similar example was encountered previously in the
context of a massive Klein-Gordon field in AdS in \cite{Andrade:2011dg}.}.
Let $h_{\pm}$ denote the solutions whose frequencies have respectively positive and negative imaginary parts. Reality of the norm requires both $(h_\pm, h_\pm) = 0$ and $(h_+, h_-) = (h_-, h_+)^*$. We can verify by
explicit calculation that this is indeed the case, with $(h_+, h_-) $ given by the right hand side of \eqref{ip l=1} (which is now
imaginary by assumption). For $M=0$, the norm vanishes identically even for $L(r_D) \neq 0$,
so the matrix of inner products in this subset is null which implies that these modes are pure gauge.
But for $M \neq 0$  the matrix of inner products for these two
modes can be diagonalized.  One of the diagonal-basis modes has positive norm while the other norm is negative, but both are non-zero for $L(r_D)\neq 0$.  So these modes are again physical. We emphasize the key role of the boundary contribution to the symplectic
structure \eqref{Omega full} in arriving at these results.

\section{Black Holes and Dirichlet Walls} \label{app:BlackHoles}

As noted in the main text, for modes that exist already in the absence of a black hole, one expects the addition of a black hole horizon on the physical side of the Dirichlet wall only to add dissipation and to make the modes more stable.  We now verify this expectation for both spherical and (for AdS) planar black holes inside respectively spherical and planar Dirichlet walls.    The $\ell_S =1$ modes were already studied in section \ref{sec:l=1 modes}, so we consider below only $\ell_S \neq 1$.

We denote the location of the black hole horizon by $r=r_H$, taken to be the appropriate solution of $f(r_H) =0$.
We thus consider modes supported in $r_H < r < r_D$ and impose ingoing boundary conditions at the horizon. In terms of the master fields, this condition takes the form \cite{Berti:2009kk}
	
\begin{equation}
 	\phi^{(I)}(r) = (r-r_H)^{ - i \omega/(4 \pi T) }( \phi^{(I)}_H + O (r-r_H)  ) , \qquad {\rm at} \quad r=r_H
 \end{equation}
where $\phi^{(I)}_H$ is a constant and $T$ is the temperature of the horizon given by
\begin{equation}
	T = \frac{f'(r_H)}{4 \pi}.
\end{equation}

\subsection{Schwarzschild}
\label{sec:schw sphere}

For $\lambda = 0$ and $M>0$ the spacetime \eqref{Schw background} is Schwarzschild, with its familiar regular event horizon at $r_H = (2M)^{1/(n-1)}$. We focus on the region $r_H < r < r_D$
between the horizon and the wall, imposing ingoing boundary conditions at the horizon: $\phi \sim (r-r_H)^{- i \omega r_H/(n-1)}$ near $r = r_H$.  As usual, stability of vector and tensor modes may be shown by adapting the argument of \cite{Ishibashi:2003ap} but we attack the scalar modes more explicitly.

\begin{figure}[h]
\center
\subfigure[][]{
\includegraphics[width=0.41\linewidth]{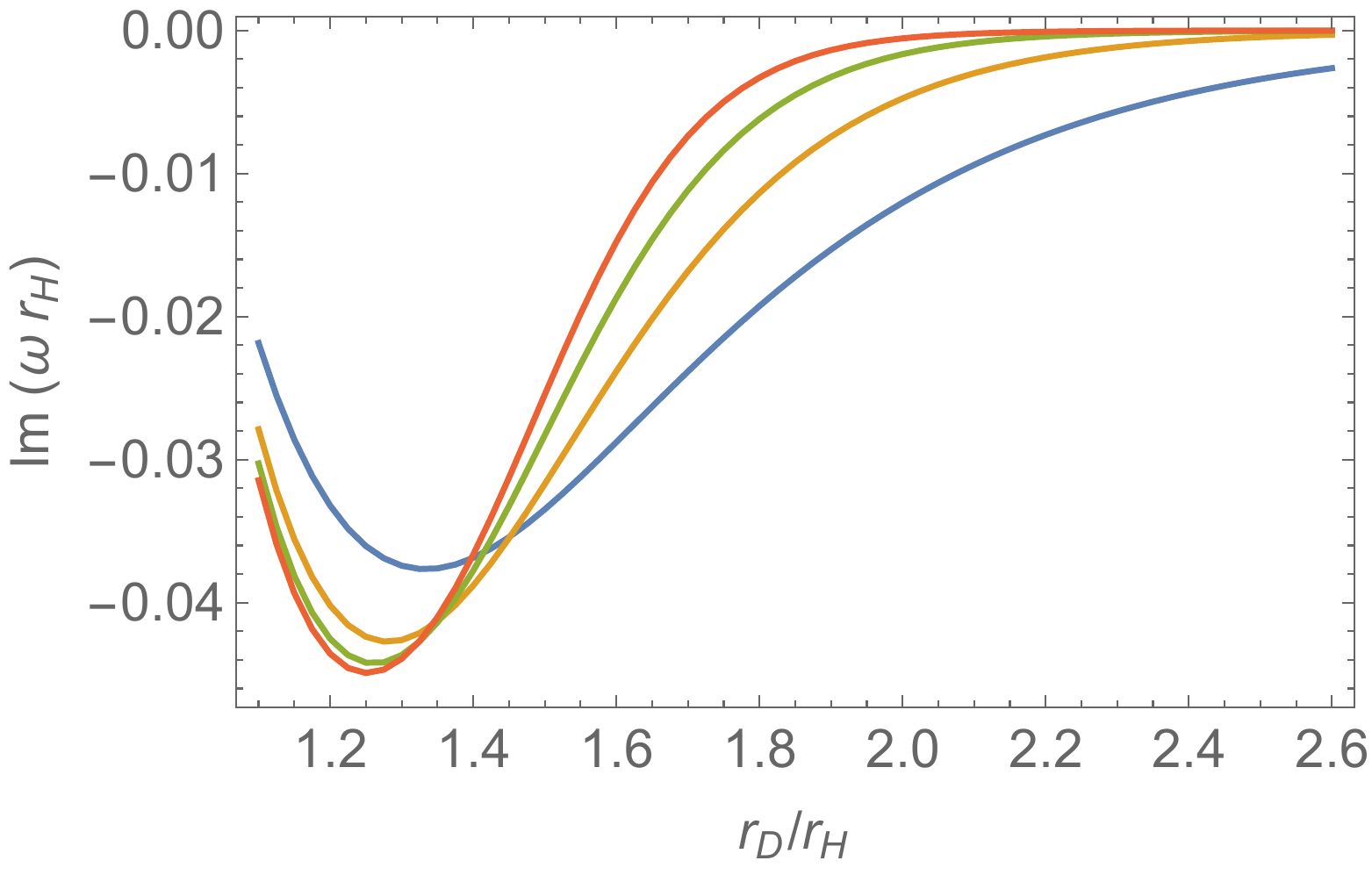}
}\qquad\qquad
\subfigure[][]{
\includegraphics[width=0.40\linewidth]{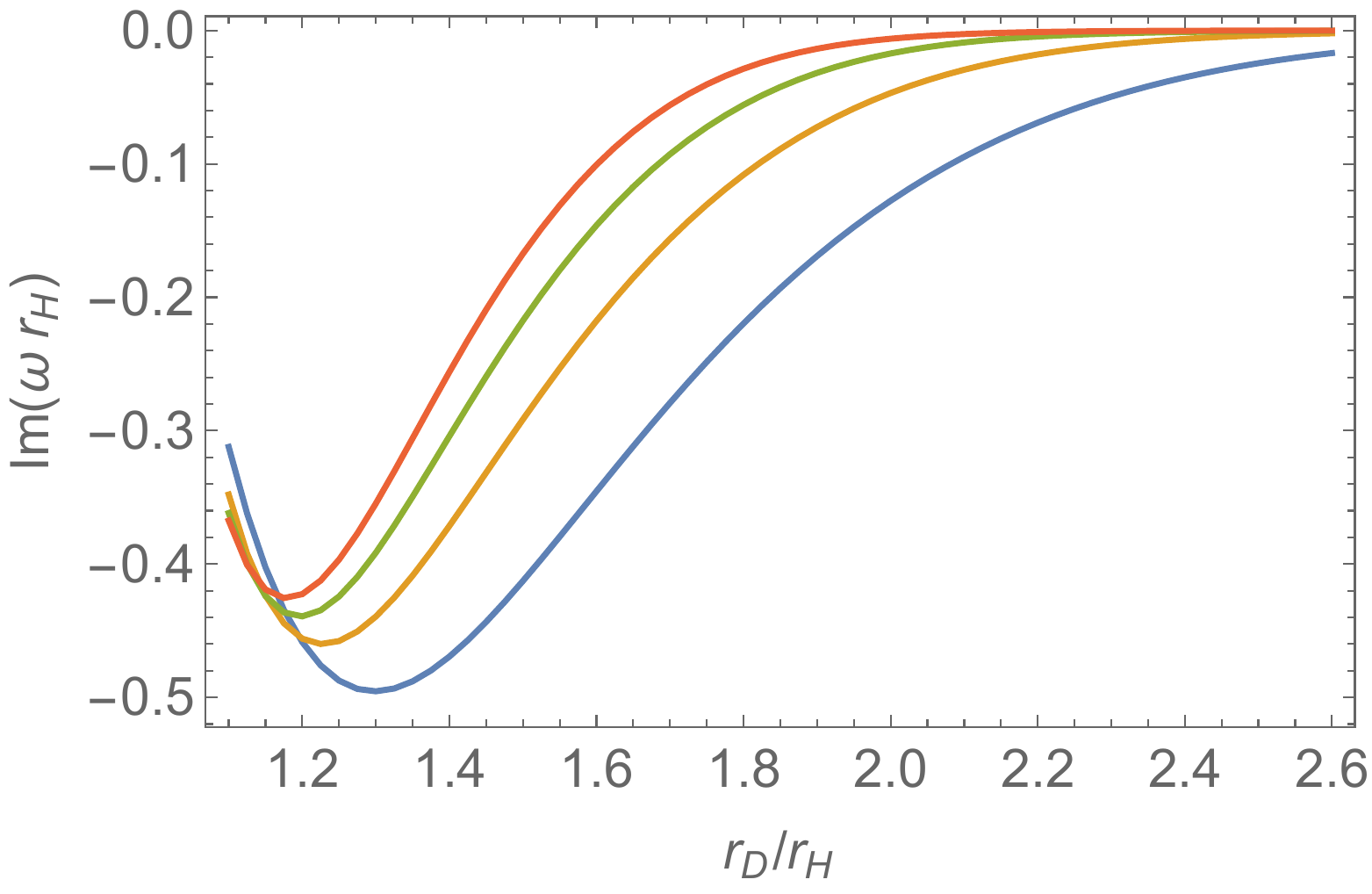}
}
\caption{Imaginary part of the lowest scalar QNM's for a Schwarzschild black hole with $\ell_S$=2, 3, 4, 5
(blue circles, yellow squares, green diamonds, red triangles)
for $n=2$ (left) and $n=5$ (right). The plots display the following expected behaviors:  i) For $r_H \ll r_D$, modes with larger $\ell_S$ experience less damping due to the stronger potential barrier that shields them from the horizon.  ii) For
$r_H$ near $r_D$ the potential barrier has comparable effects on the $\ell_S =2,3,4,5$ modes.  In this regime, the growth of positive gradient energy with $\ell_S$ makes the $\ell_S=5$ mode the most stable of the ones shown.}
\label{Sch QNM 1}
\end{figure}

For non-vanishing $M$ the equation of motion is solved by Heun functions, about which rather little is known.  We find it most convenient to simply solve numerically for these functions and the associated scalar quasi-normal modes (QNM) by discretizing the wave equation using a Chebyschev grid and solving the resulting
discrete eigenvalue problem. In this way, we have determined the first five QNM  for $2 \le n \le 5$,
$2 \le \ell_I \le 10 $, $0.1 \le r_D/r_H \le 10$ finding no unstable modes, see figure
\ref{Sch QNM 1}. For each set of quantum numbers the modes decay at late times and may thus be said to be more stable than for $M=0$ (where the modes oscillate but neither grow nor decay).

\subsection{Schwarzschild dS}

\begin{figure}[h]
\center
\subfigure[][]{
\includegraphics[width=0.4\linewidth]{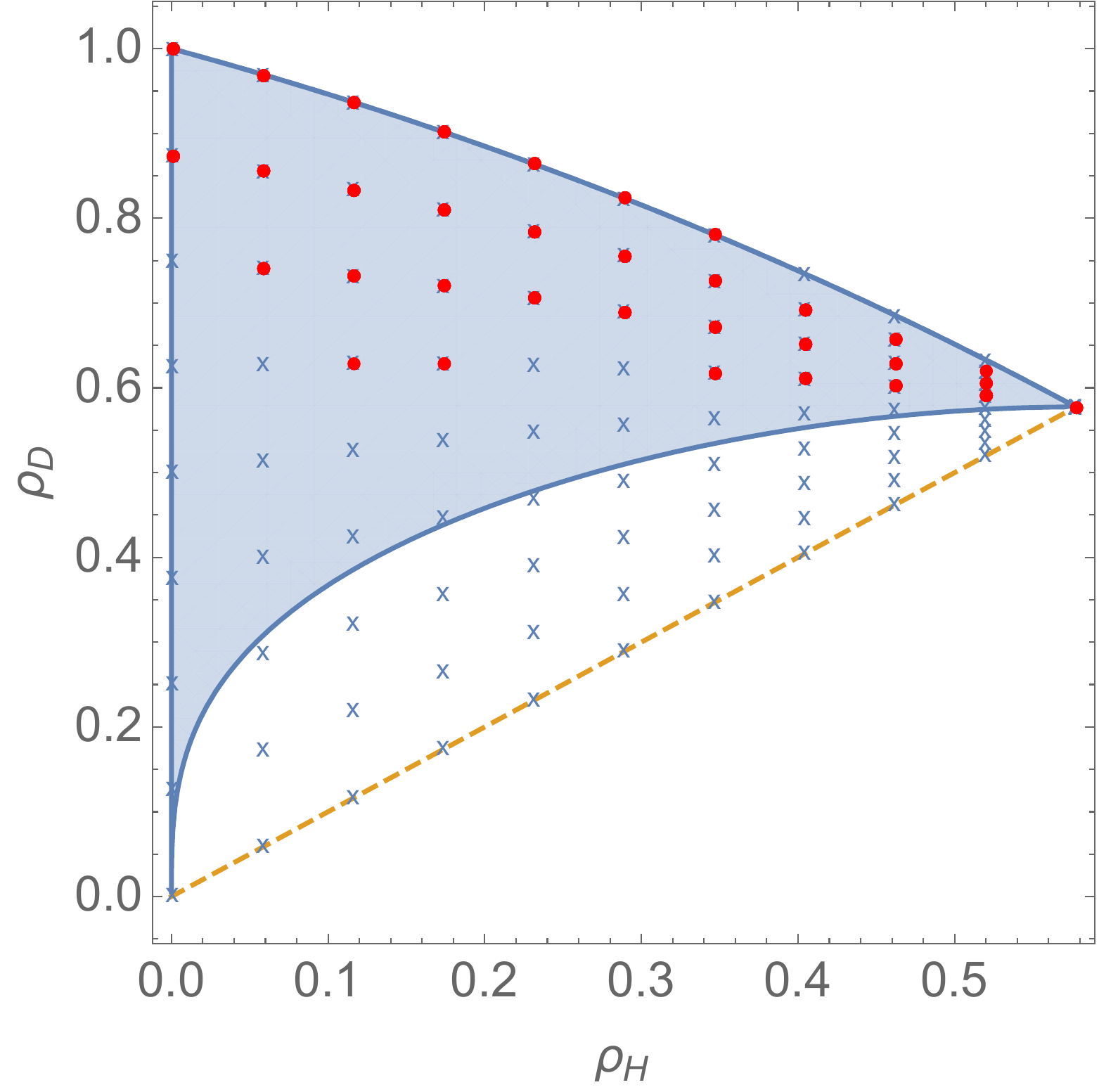}
}\qquad\qquad
\subfigure[][]{
\includegraphics[width=0.40\linewidth]{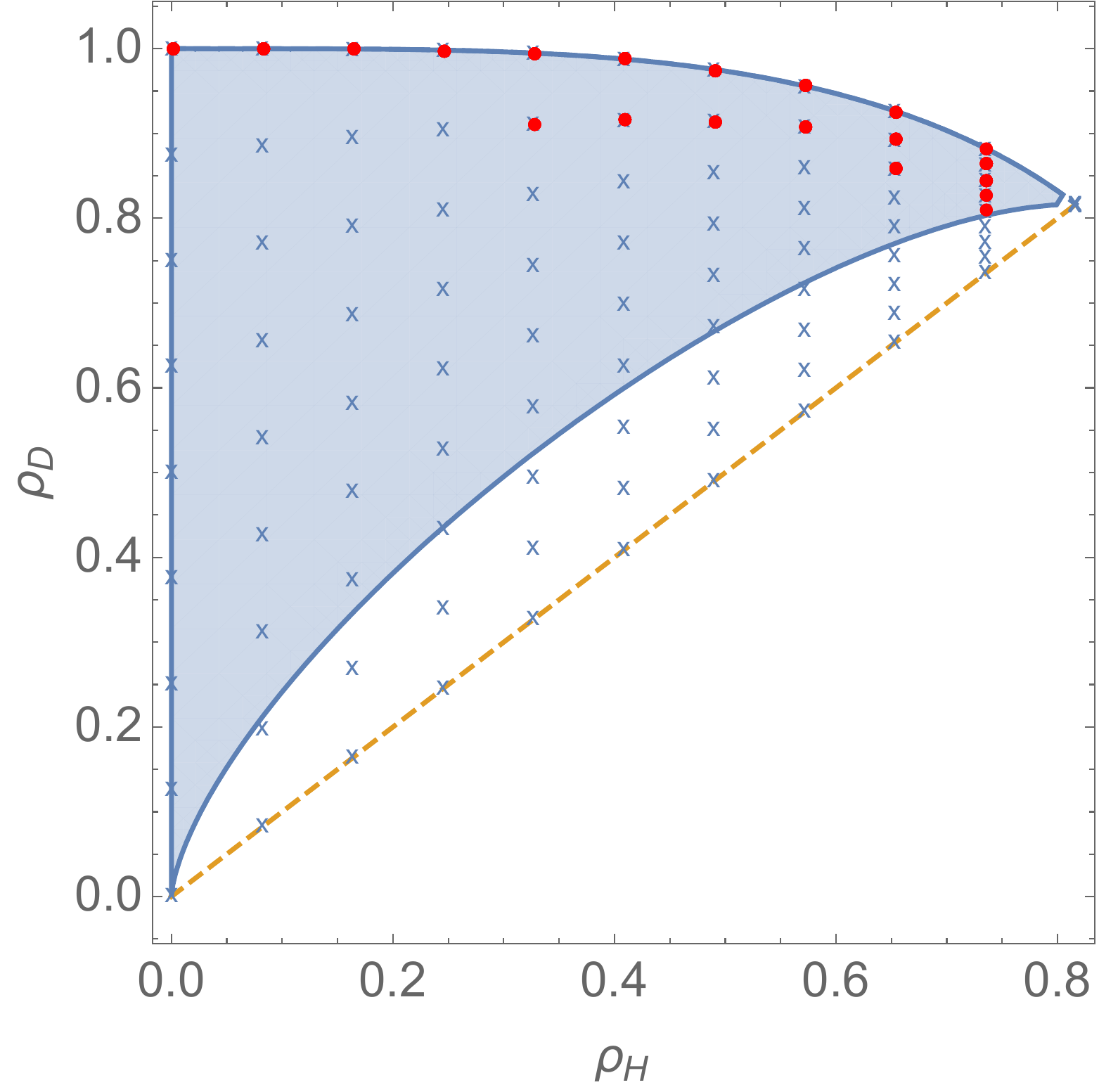}
}
\caption{Stability region of Schwarzschild AdS black holes for $n=2$ (left) and $n=5$ (right). In the shaded region
the $\ell_S = 1$ diffeo modes have imaginary frequencies are thus unstable, i.e. $f'(\rho_D) < 0$. The red circles correspond to
positions in the parameter space in which we have found unstable QNM for $2 \le l_S \le 10$. The crosses indicate other points at which we have performed our numerical search, and where no instabilities were found.  The figures suggest that stability of the $\ell_S=1$ modes also implies stability of all modes with $\ell_S \ge 2$.}
\label{schw dS QNM}
\end{figure}

Spherical Schwarzschild dS black holes, corresponding to $K=1$, $M>0$, $\lambda > 0$ in
\eqref{Schw background}, can be studied in much the same manner. This geometry exhibits two horizons: the event horizon, $r_H$, and the cosmological horizon $r_{cosm}$, with
$r_H < r_D$.  We restrict ourselves to $r_H < r_D < r_{cosm}$ and consider modes in the region $r_H < r < r_D$. The temperature of the black hole horizon is
\begin{equation}
	T_{{\rm Sch, dS}} = 	\frac{1}{4\pi r_H}\left( n-1 - (n+1)\frac{r_H^2}{L_{dS}^2} \right) >0.
\end{equation}
Ingoing boundary conditions at the
horizon require $\phi \approx (r-r_H)^{- i \omega/(4 \pi T_{{\rm Sch, dS}})}$ at
$r=r_H$. We solve the for the spectra of QNM numerically as described in section \ref{sec:schw sphere} for $2 \le n \le 5$
in units with $L_{dS} = 1$.
We find no evidence of instabilities in the tensor or vector sectors\footnote{Indeed, stability of the tensor sector can again be shown as in \cite{Horowitz:1999jd}.}, but the generic lowest lying modes in the
scalar sector become unstable for sufficiently large cavities, see figure \ref{schw dS QNM}.
This is expected from the $M=0$ results of section \ref{sec:dsSphere} which found unstable scalar modes but only stable vector and tensor modes.  It seems that large black holes introduce large dissipation that completely removes the $M=0$ scalar instabilities.
In addition to the
unstable generic scalar modes, the dS black holes can suffer from instabilities corresponding to the $\ell_S = 1$ modes. These
occur in a large region of the parameter space $(\rho_H, \rho_D)$ characterized by $f'(\rho_D) <0$; see figure \ref{schw dS QNM}.  \\

\subsection{Spherical Schwarzschild AdS}

Spherical Schwarzschild AdS black holes are described by taking $K=1$, $M>0$, $\lambda<0$
in \eqref{Schw background}.
The black hole hoirzon lies at the unique positive solution $r_H$ of $f(r_H) = 0$.  As for $\lambda = 0$, we consider
modes in the region $r_H < r < r_D$. Ingoing boundary conditions at the horizon read $\phi \approx (r-r_H)^{-i \omega/(4\pi T) }$
where
\begin{equation}
	T_{{\rm Sch, AdS}} = 	\frac{1}{4\pi r_H}\left( n-1 + (n+1)\frac{r_H^2}{L_{AdS}^2} \right).
\end{equation}

Once again, we solve for the QNM spectrum numerically by discretezing the eigenvalue
problem given by the master field equation with the corresponding boundary conditions. We perform the numerics using the dimensionless
radial variable $\rho = r/L_{AdS}$, and denote $\rho_H = r_H/L_{AdS}$ and $\rho_D= r_D/L_{AdS}$ the locations of the horizon
and Dirichlet wall in units of the AdS radius.

\begin{figure}[h]
\center
\includegraphics[width=0.8\linewidth]{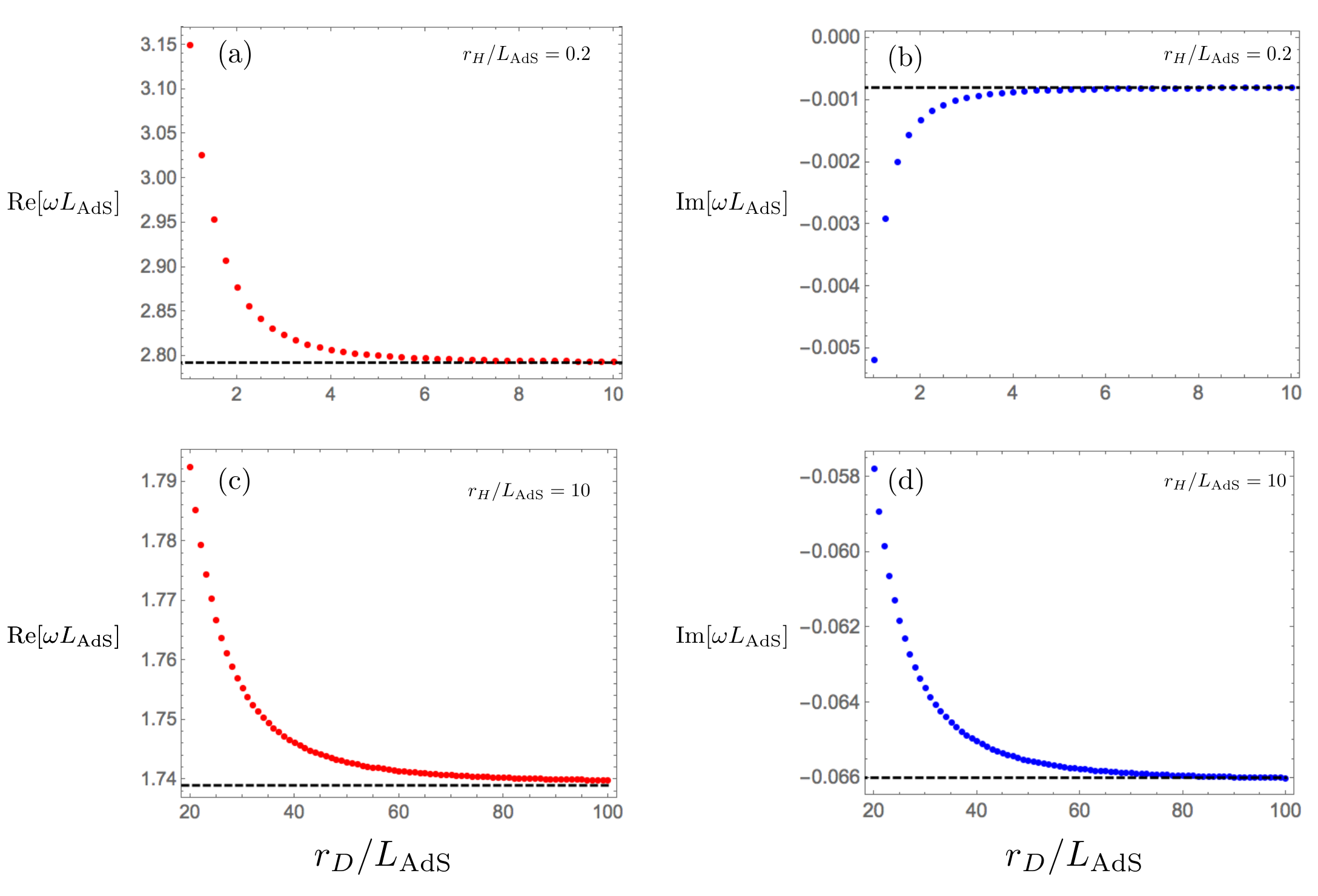}
\caption{
Real (a,c) and imaginary (b,d) parts of the frequency of the least stable scalar mode with $\ell_S=2$ of a Schwarzschild AdS black hole in a spherical cavity.  We vary the ratio $r_D/L_\text{AdS}$ for a small (a,b) and a large (c,d) black hole.  The black dashed line are the results for AdS Schwarzschild without a Dirichlet wall (taken from table 2 of \cite{Michalogiorgakis:2006jc}).}
\label{small schw AdS QNM}
\end{figure}

The tensor and vector modes can again be shown to be stable by adapting \cite{Ishibashi:2003ap}.  We thus consider only the scalar modes in
detail. For $2 \le n \le 5$  we have computed the first four QNM modes sampling over
$0.1 \le \rho_H \le \rho_D \le 10$, $2 \le \ell_S \le 15$ , finding no instabilities. In figure~\ref{small schw AdS QNM} we plot the frequency  of the least stable scalar mode with $\ell_S = 2$.  We find that the imaginary part of the frequency remains negative and approaches the asymptotically AdS value in the limit $r_D/L_\text{AdS} \gg 1$.

\subsection{Planar Schwarzschild AdS}

\begin{figure}[h]
\center
\subfigure[][]{
\includegraphics[width=0.4\linewidth]{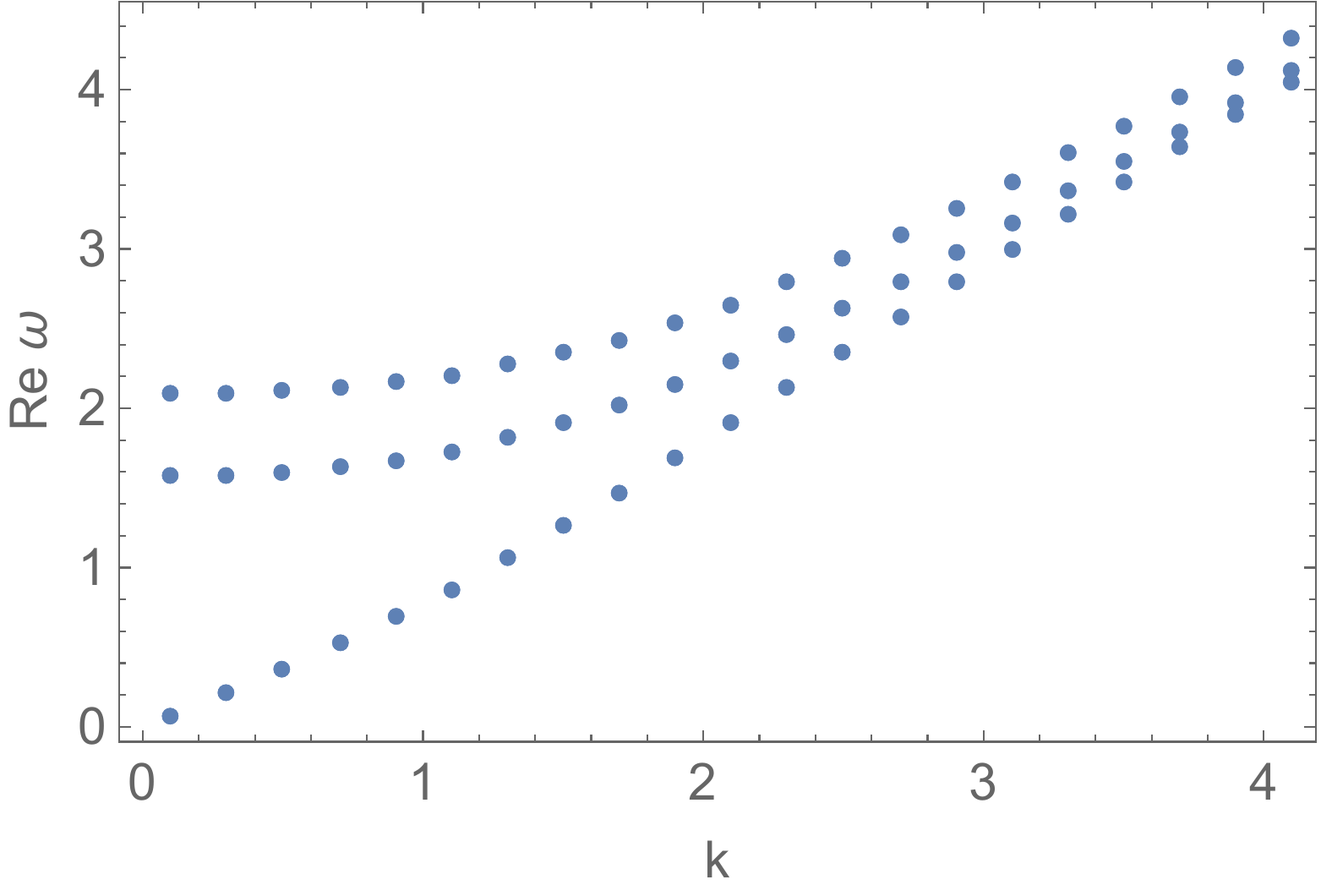}
}\qquad\qquad
\subfigure[][]{
\includegraphics[width=0.40\linewidth]{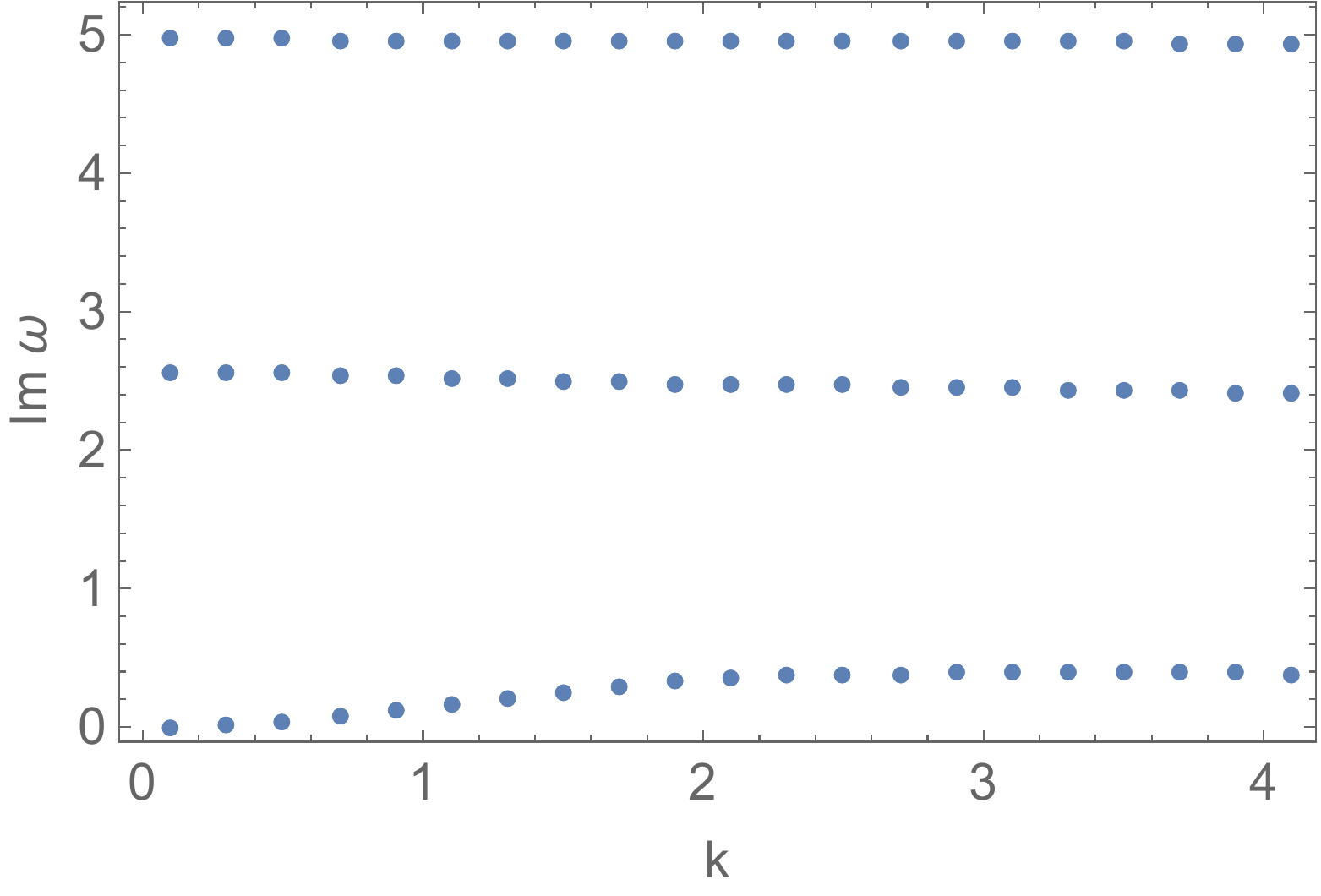}
}
\caption{Real and imaginary part of the first scalar QNM for a planar $AdS_4$ Schwarzschild black hole
inside a cavity of size $y_D = 3$ as a function of $\hat k$. The lowest lying mode is hydrodynamic.}
\label{schw AdS QNM 1}
\end{figure}

\begin{figure}[h]
\center
\subfigure[][]{
\includegraphics[width=0.4\linewidth]{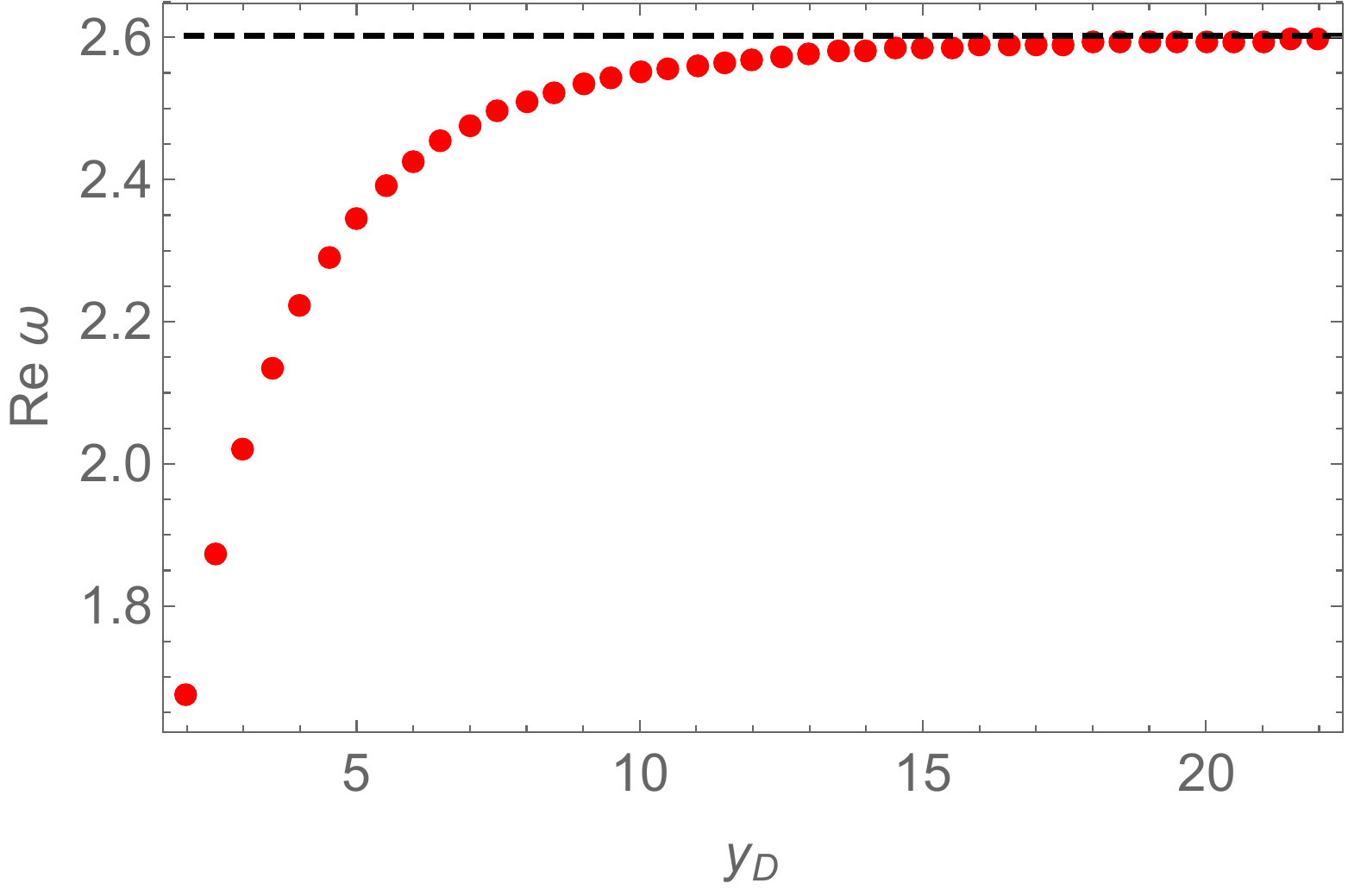}
}\qquad\qquad
\subfigure[][]{
\includegraphics[width=0.42\linewidth]{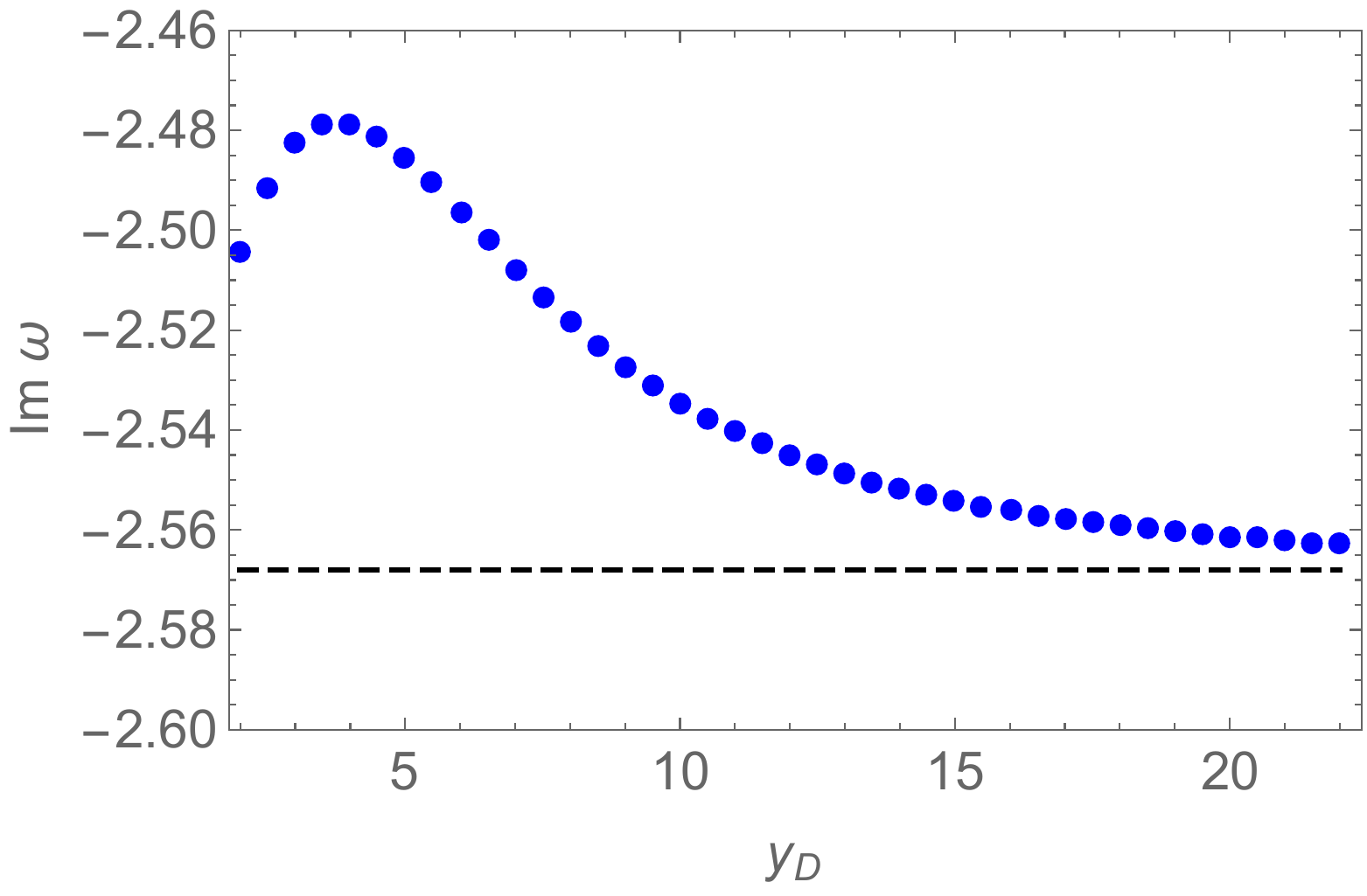}
}
\caption{Real and imaginary part of the first $k=2$ scalar 
non-hydrodynamic QNM for a planar $AdS_4$ Schwarzschild black hole
as a function of $y_D$. The dashed line corresponds to a QNM computed for the full spacetime, obtained from table 2
in \cite{Miranda:2008vb}.}
\label{schw AdS QNM 2}
\end{figure}

We finally consider fluctuations around a planar Schwarzschild AdS black
hole inside a flat cavity at $r = r_D$. We restrict to the interior region $r_H < r < r_D$ and set $L_{AdS} =1$.
The event horizon is then located at $r_H = (2M)^{1/(n+1)}$.  We impose
ingoing boundary conditions at the horizon, which require the solutions to behave as
$\phi \approx (r-r_H)^{-i \omega/[r_H (n+1)]}$ near $r = r_H$.
As usual, adapting the argument of \cite{Ishibashi:2003ap} shows the tensor and vector modes to be stable,
so we need only solve for the QNM spectra of the scalar modes.  Our numerical method
is to discretize the wave equation and solve the resulting linear eigenvalue equation.
We have obtained numerically the first four scalar QNMs for $2 \le n \le 5$ and a wide range of values of
the spatial momentum $\hat k_S = k_S/r_H$ and locations of the wall $y_D = r_D/r_H$  finding that all the modes are stable, see
figures \ref{schw AdS QNM 1} and \ref{schw AdS QNM 2} for sample plots. For large values of $y_D$, our results coincide with
those obtained in the full range of the radial coordinate, computed previously in \cite{Miranda:2008vb} for $n=2$.

\clearpage

\bibliographystyle{JHEP-2}
\bibliography{DirichletStability}

\end{document}